\documentclass{aa}		
\usepackage[english]{babel}
\usepackage[utf8]{inputenc}
\usepackage{amsmath,amsfonts,amssymb,enumerate}
\usepackage{txfonts,graphics,graphicx,epstopdf,float,lscape,longtable,dcolumn,footnote,subcaption,caption,color,url,hyperref,listings,esvect}
\usepackage{rotating,afterpage}
\usepackage{ulem}
\usepackage[symbol]{footmisc}

\usepackage{soul,xcolor}

\setstcolor{red}

\newcommand*\xbar[1]{{\hbox{\vbox{\hrule height 0.5pt \kern0.5ex \hbox{\kern-0.1em \ensuremath{#1} \kern-0.1em}}}}}

\usepackage{natbib}
	\bibpunct{(}{)}{;}{a}{}{,} 

\defcitealias{IbanezMejia2016}{Paper I}
\defcitealias{IbanezMejia2017}{Paper II}
\defcitealias{Chira2018}{Paper III}

\def\apjl{{Astrophys. J. Lett.} }

\def\aap{{Astron. Astrophys.} }

\def\apjs {{Astrophys. J. Suppl.}}

\def\H    {\hbox{\rm H}}     

\title{How do velocity structure functions trace gas dynamics in simulated molecular clouds?}
\author{
	R.-A.~Chira\inst{\ref{mpia}} \and
	J.~C.~Ib\'a\~{n}ez-Mej\'{\i}a\inst{\ref{koeln},\ref{mpe}} \and 
	M.-M.~Mac~Low\inst{\ref{amnh},\ref{ita}, \ref{fi}} \and
	Th.~Henning\inst{\ref{mpia}}
  }
\institute{
	Max-Planck-Institut f\"ur Astronomie, K\"onigstuhl 17, 69117 Heidelberg, Germany\\ \email{roxana-adela.chira@alumni.uni-heidelberg.de}\label{mpia}
	\and I.\ Physikalisches Institut, Universit\"at zu K\"oln,
        Z\"ulpicher Straße 77, 50937 K\"oln, Germany\\ \email{ibanez@ph1.uni-koeln.de}\label{koeln}
        \and Max-Planck-Institut f\"ur Extraterrestrische Physik,
          Giessenbachstrasse 1, 85748 Garching, Germany\label{mpe}
	\and Dept.\ of Astrophysics, American Museum of Natural History, 79th St.\ at Central Park West, New York, NY 10024, USA\\ \email{mordecai@amnh.org}\label{amnh}
	\and Zentrum f\"ur Astronomie, Institut f\"ur Theoretische
        Astrophysik, Universit\"at Heidelberg, Albert-Ueberle-Str.\ 2, 69120 Heidelberg, Germany\label{ita}
    \and Center for Computational Astrophysics, Flatiron Institute, 162 Fifth Ave, New York, NY 10010, USA\label{fi}
}

\date{accepted for publication on August 9, 2019}

\abstract
	{ 
        Supersonic disordered flows accompany the formation and evolution of molecular clouds (MCs). 
        It has been argued that this is turbulence that can support against gravitational collapse and form hierarchical sub-structures. 
	}
	{ 
    	We investigate the time evolution of simulated MCs to investigate:
        What physical process dominates the driving of turbulent flows?
        How can these flows be characterised?  
        Are they consistent with uniform turbulence or gravitational collapse?
        Do the simulated flows agree with observations?
	} 
	{ 
    	We analyse three MCs that have formed self-consistently within kiloparsec-scale numerical simulations of the interstellar medium (ISM). 
        The simulated ISM evolves under the influence of physical processes including self-gravity, stratification, magnetic fields, supernova-driven turbulence, and radiative heating and cooling. 
        We characterise the flows using velocity structure functions
        (VSFs) with and without density weighting or a density cutoff, and computed in one or three dimensions. 
        However, we do not include optical depth effects that can hide motions in the densest gas, limiting comparison of our results with observations.
	}
	{ 
	    In regions with sufficient resolution, the density-weighted VSFs initially appear to follow the expectations for uniform turbulence, with a first-order power-law exponent consistent with Larson's size-velocity relationship. 
        SN blast wave impacts on MCs produce short-lived coherent motions at large scales, increasing the scaling exponents for a crossing time. Gravitational contraction drives small-scale motions, producing scaling coefficients that drop or even turn negative as small scales become dominant.
        Removing the density weighting eliminates this effect as that emphasises the diffuse ISM.
	}
	{ 
        We conclude that two different effects coincidentally reproduce Larson's size velocity relationship. 
        Initially, uniform turbulence dominates, so the energy cascade produces VSFs consistent with Larson's relationship. 
        Later, contraction dominates, the density-weighted VSFs become much shallower or even inverted, but the relationship of the global average velocity dispersion of the MCs to their radius
        follows Larson's relationship, reflecting virial equilibrium or free-fall collapse.
        The injection of energy by shocks is visible in the VSFs, but decays within a crossing time.
       }

	\keywords{Turbulence -- ISM: kinematics and dynamics -- ISM: structure -- ISM: clouds}

\begin{document}
	\maketitle

\section{Introduction}\label{intro}

It has long been known that star formation preferentially occurs within molecular clouds (MCs). 
However, the physics of the star formation process is still not completely understood.
It is obvious that gravity is the key factor for star formation as it drives collapse motions and operates on all scales.
However, one needs additional processes that stabilise the gas or terminate star formation quickly in order to explain the low star formation efficiencies observed in MCs. 
Although there are many processes that act at the different scales of MCs, turbulent support has often been argued to be the best candidate for this task.

In the literature, turbulence plays an ambiguous role in the context of star formation. 
In most of the cases, turbulence is expected to stabilise MCs on large scales \citep{Fleck1980,McKee1992,MacLow2003}, while feedback processes and shear motions heavily destabilise or even disrupt cloud-like structures \citep{Tan2013,Miyamoto2014}. 
However, it remains unclear whether there are particular mechanisms that dominate the driving of turbulence within MCs, as every process is supposed to be traced by typical features in the observables.
Yet, these features are either not seen or are too ambiguous to clearly reflect the dominant driving mode.
For example, turbulence that is driven by large-scale velocity dispersions during global collapse \citep{Ballesteros2011a,Ballesteros2011b,Hartmann2012} produces P-Cygni line profiles that have not yet been observed on scales of entire MCs. 
Internal feedback, on the contrary, seems more promising as it drives turbulence from small to large scales \citep{Dekel2013,Krumholz2014}.
Observations, though, demonstrate that the required driving sources need to act on scales of entire clouds; which typical feedback, such as radiation, winds, jets, or supernovae (SNe), cannot achieve \citep{Heyer2004,Brunt2009,Brunt2013}.

There have been many theoretical studies that have examined the nature and origin of turbulence within the various phases of the interstellar medium \citep[ISM;][and references within]{MacLow2004}. 
The most established characterisation of turbulence in general was introduced by \citet{Kolmogorov1941} who investigated fully developed, incompressible turbulence driven on scales larger than the object of interest, and dissipating on scales much smaller than those of interest.
In the scope of this paper this object is a single MC. 
MCs are highly compressible, though.
Only a few analytical studies have treated this case.
\citet{She1994} and \citet{Boldyrev2002}, for example, generalise and extend the predicted scaling of the decay of turbulence to supersonic turbulence.
\citet{Galtier2011} and \citet{Banerjee2013} provide an analytic description of the scaling of mass-weighted structure functions.

\citet{Larson1981} found a relation between the linewidth $\sigma$ and the effective radius $R$ of MCs.
Subsequent investigators have settled on the form of the relation being \citep{Solomon1987,Falgarone2009,Heyer2009}
\begin{equation} \label{eq:larson}
    \sigma \propto R^{1/2}.
\end{equation}
\citet{Goodman1998} showed that analysis techniques used to study this relation could be distinguished by whether they studied single or multiple clouds using single or multiple tracer species.
Explanations for this relation have relied on either turbulent cascades \citep{Larson1981,Kritsuk2013,Kritsuk2015,Gnedin2015,Padoan2016}, or the action of self-gravity \citep{Elmegreen1993,Vazquez2006,Elmegreen2007,Heyer2009,Ballesteros2011}.

These can potentially be distinguished by examining the velocity structure function.
\citet{Kritsuk2013} carefully reviews the argument for Larson's size-velocity relation depending on the turbulent cascade. 
In short, in an energy cascade typical for turbulence, the second-order structure function has a lag dependence $\ell^{\zeta(2)}$ with $\zeta(2) \simeq 1/2$. 
In \citet[hereafter \citetalias{IbanezMejia2016}]{IbanezMejia2016} the authors argued that uniform driven turbulence was unable to explain the observed relation in a heterogeneous interstellar medium, but that the relation could be naturally explained by hierarchical gravitational collapse.

In this paper, we examine the velocity structure functions of three MCs that formed self-consistently from SN-driven turbulence in the simulations by \citetalias{IbanezMejia2016} and \citet[][hereafter \citetalias{IbanezMejia2017}]{IbanezMejia2017}.

and study how the turbulence within the clouds' gas evolves.
The key questions we address are the following:
What dominates the turbulence within the simulated MCs? 
Does the observed linewidth-size relation arise from the turbulent flow?
How can structure functions inform us about the evolutionary state of MCs and the relative importance of large-scale turbulence, discrete blast waves, and gravitational collapse?

In Sect.~\ref{methods}, we introduce the simulated clouds in the context of the underlying physics involved in the simulations.
Furthermore, we describe the theoretical basics of velocity structure functions.
Sect.~\ref{results} demonstrates that the velocity structure function is a useful tool to characterise the dominant driving mechanisms of turbulence in MCs and can be applied to both simulated and observed data. 
We examine the influence of using one-dimensional velocity measurements, different Jeans refinement levels, density thresholds, and density weighting on the applicability of the velocity structure function and the results obtained with it in Sect.~\ref{discussion}.  
At the end of that section, we will also compare our results to observational studies.
We summarise our findings and conclusions in Sect.~\ref{conclusions}.
The simulation data and the scripts that this work is based on are published in the Digital Repository of the American Museum of Natural History \citep{Chira2018b}.

\section{Methods}\label{methods}

\subsection{Cloud models}\label{methods:clouds}

The analysis in this paper is based on a sample of three MCs identified within a three-dimensional (3D), magnetohydrodynamical, adaptive mesh refinement simulation using the FLASH code \citep{Fryxell2000}.  
\citetalias{IbanezMejia2016} and \citetalias{IbanezMejia2017}, as well as \citet[\citetalias{Chira2018} hereafter]{Chira2018}, describe the simulations and the clouds in more detail. 
We summarise the most relevant properties. 

The numerical simulation models a $1\times1\times40$~kpc$^3$ section of the multi-phase, turbulent ISM of a disk galaxy, where dense structures form self-consistently in convergent, turbulent flows \citepalias{IbanezMejia2016}.  
The model includes gravity---a background galactic-disk potential accounting for a stellar component and a dark matter halo, as well as self-gravity turned on after 250~Myr of simulated time---SN-driven turbulence, photoelectric heating and radiative cooling, and magnetic fields. 
Although hundreds of dense clouds form within the simulated volume, \citetalias{IbanezMejia2017} focused on three clouds, which were re-simulated at a much higher spatial resolution.
The internal structures of the MCs are resolved using adaptive mesh refinement, focusing grid resolution on dense regions where Jeans unstable structures must be resolved with a minimum of 4 cells ($\lambda_J > 4~\Delta x_{\rm min}$).
For a maximum resolution of $\Delta x = 0.1$~pc, the corresponding maximum resolved density is $8~\times 10^3$~cm$^{-3}$ for gas at a temperature of 10~K \citepalias[e.g.][Eq.~15]{IbanezMejia2017}. 
We define MCs as regions above a fixed number density threshold with fiducial value $n_{\mathrm cloud}$~=~100~cm$^{-3}$.
We chose this threshold as it roughly corresponds to the density when CO becomes detectable.
The MCs have initial masses at the onset of self-gravity of $3~\times 10^3$, $4~\times 10^3$, and $8~\times 10^3$~M$_{\odot}$ and are denoted as \texttt{M3}, \texttt{M4}, and \texttt{M8}, respectively, hereafter. 
In this paper, we use the data within (40~pc)$^{3}$ subregions centred on the high-resolution clouds' centres of mass, which we map to a uniform grid of 0.1 pc zones for analysis.
For illustrations of the morphologies of the three clouds we refer to Fig.~1 of \citetalias{Chira2018}.

It is important to point out that the clouds are embedded within a complex turbulent environment, gaining and losing mass as they evolve.
\citetalias{IbanezMejia2017} described the time evolution of the properties of all three clouds in detail, in particular, mass, size, velocity dispersion, and accretion rates, in the context of MC formation and evolution within a galactic environment.
\citetalias{Chira2018} studied the properties, evolution, and fragmentation of filaments that self-consistently condense within these clouds. 
We paid particular attention to the accuracy of typical stability criteria for filaments, comparing the results to the theoretical predictions, showing that simplified analytic models do not capture the complexity of fragmentation due to their simplifying assumptions.

\subsection{Velocity Structure Function}\label{methods:vsf}

In this paper, we probe the power distribution of turbulence throughout the entire simulated MCs by using the velocity structure function (VSF).
The VSF is a two-point correlation function,
\begin{equation}
	{S}_p (\ell) = \langle \, |\Delta \vec{v}|^p  \, \rangle
	\label{equ:method:def_vsf}
\end{equation}
that measures the mean velocity difference 
\begin{equation}
    \Delta \vec{v} (\vec{\ell}) = \vec{v}(\vec{x}+\vec{\ell}) - \vec{v}(\vec{x})
\end{equation} 
between two points $\vec{x}$ and $\vec{x}+\vec{\ell}$, with $\vec{\ell}$ being the direction vector pointing from the first to the second point.
The VSF $S_p$ is usually reported as a function of lag distance, $\ell = |\vec{\ell}|$, between the correlated points.
The coherent velocity differences measured by the VSF can be produced by both the energy cascade expected in turbulent flows, and by coherent motions such as collapse, rotation, or blast waves.
Those patterns become more prominent the higher the value of the power $p$ is \citep{Heyer2004}.

For fully developed, homogeneous, isotropic, turbulence the VSF is well-described by a power-law relation \citep{Kolmogorov1941,She1994,Boldyrev2002}:
\begin{equation}
	\mathit{S}_p (\ell) \propto \ell^{\zeta(p)}.
	\label{equ:method:propto_zeta}
\end{equation}

\citet{Kolmogorov1941} predicts that the third-order exponent, $\zeta(3)$, is equal to unity for an incompressible flow.
As a consequence the kinetic energy decays with $E_{\mathrm{kin}}(k) \propto k^{-\frac{5}{3}}$, with $k = \frac{2 \pi}{\ell}$ being the wavenumber of the turbulence mode.
For a supersonic flow, however, $\zeta(3) >1$ is expected.
Based on \citeauthor{Kolmogorov1941}'s work, \citet{She1994} and \citet{Boldyrev2002} extended and generalised the analysis and predicted the following intermittency corrections to \citeauthor{Kolmogorov1941}'s scaling law.
For incompressible turbulence with filamentary dissipative structures \citet{She1994} predict that
the VSFs scale with power law index
\begin{equation}
	\zeta(p) = \frac{p}{9} + 2 \left[ 1 - \left( \frac{2}{3} \right)^{\frac{p}{3}} \right] ,
	\label{equ:method:she}
\end{equation}
while supersonic flows with sheet-like dissipative structures are predicted to scale with \citep{Boldyrev2002}
\begin{equation}
	\zeta(p) = \frac{p}{9} + 1 - \left( \frac{1}{3} \right)^{\frac{p}{3}}.
	\label{equ:method:boldyrev}
\end{equation}
\noindent
Note that both equations return a value of $\zeta(3) =1$, but this is only an exact result for the \citeauthor{She1994} model, while it is a result of normalisation in the case of \citeauthor{Boldyrev2002}. 

In the case of compressible turbulence, the energy cascade can no longer be expressed in terms of a pure velocity difference because density fluctuations become important.
Turbulence should then show a cascade in some density-weighted VSF analogous to the incompressible case.
\citet{Padoan2016a} defined a density-weighted VSF to attempt to capture this process, which we use in our subsequent analysis
\begin{equation}
	{S}_p (\ell) = \frac{\langle \, \rho(\vec{x}) \rho(\vec{x}+\vec{\ell}) \, |\Delta \vec{v}|^p  \, \rangle}{\langle  \, \rho(\vec{x}) \rho(\vec{x}+\vec{\ell}) \, \rangle}.
	\label{equ:method:def_vsf_dw}
\end{equation}
Alternatives have been proposed by \citet{Kritsuk2013a} based on an analysis of the equations of compressible flow that should be explored in future work.

In many cases a three-dimensional computation of the VSF cannot be performed because of the observational constraint that only line-of-sight velocities are available.
We therefore compare our three-dimensional (3D) results to one-dimensional (1D), density-weighted VSFs
\begin{equation}
	\mathit{S}_{p,\mathrm{1D}} (\ell) = \frac{\langle \, \rho(\vec{x}) \rho(\vec{x}+\vec{\ell}) \, |\Delta 
        \vec{v} \cdot \vec{e}_i|^p  \, \rangle}{\langle  \, \rho(\vec{x}) \rho(\vec{x}+\vec{\ell}) \, \rangle} ,
	\label{equ:method:def_vsf_1d}
\end{equation}
with $\vec{e}_i$ representing the unit vector along the $i$~=~$x$-, $y$-, or $z$-axis.

\citet{Benzi1993} introduced the principle of extended self-similarity.
It proposes that there is a constant relationship between the scaling
exponents of different orders at all lag scales so that $\zeta$ can be measured from $S_p/S_3$, which typically gives a clearer power-law behaviour.
The self-similarity parameter is defined as,
\begin{equation}
	Z(p) = \frac{\zeta(p)}{\zeta(3)}.
	\label{equ:method:z_def}
\end{equation} 
\noindent
As mentioned before, both Eq.~(\ref{equ:method:she}) and~(\ref{equ:method:boldyrev}) return values of $\zeta(3)$~=~1.
Therefore, those equations also offer predictions for $Z(p)$.

For the discussion below, we measure $\zeta$(p) by fitting a power-law, given by
\begin{equation}
	\log_{10}\left[ S_p(\ell) \right] = \log_{10}\left(A\right) + \zeta(p) \, \log_{10}(\ell) ,
	\label{equ:method:fitting}
\end{equation}
with $A$ being the proportionality factor of the power-law to the simulated measurements.
We choose the smallest lag of the fitting range to be equal to eight zones, sufficiently large to ensure that our fit does not include the numerical dissipation range.
For more details of the fitting procedure we refer to Appendix~\ref{appFitting}.

We follow observational practice and reduce the computational effort of this study by generally focusing on clouds defined by a density threshold.
However, \citetalias{IbanezMejia2017} shows that there is usually no sharp increase in density between the ISM and the clouds. 
Instead, the gas becomes continuously denser towards the centres of mass within the clouds. 
Consequently, our use of a density threshold is a somewhat artificial boundary between the clouds and the ISM. Observationally, however, introducing a column density (or intensity) threshold is unavoidable, be it due to technical limitations (e.g., detector sensitivity) or the nature of the underlying physical processes (for example, excitation rates, or critical densities).
Therefore, we also study how a density threshold influences the VSF and its evolution.

At our fiducial density threshold, we actually consider only $\leq$1.5\% of the volume in the high resolution cube.
To understand the influence of this limitation we set up a test scenario (see Sect.~\ref{results:densthres}) by removing the density threshold (setting n$_\mathrm{cloud}$~=~0~cm$^{-3}$) that results in analysing the entire data cube.
Details of the method for computing the VSFs in these two cases are given in Appendix~\ref{appFitting}.

As in the case without a density threshold it would be too computationally expensive to compute all lags to all zones.
Thus, we randomly choose a set of 5\% of the total number of zones as reference points and only compute relative velocities from the entire cube to these zones.
By choosing the starting points randomly we ensure that all parts of the cubes are considered. 
As a consequence, there is only a small likelihood (5\%~$\times$~1.5\%~=~0.075\%) that any given zone chosen will be within the cloud.
Therefore, we emphasise that it is likely that the two subsamples (no density threshold and cloud-only) do not have a common subset of starting vectors.
Nevertheless, the random sample still includes $>4 \times 10^3$ zones in the cloud, so the sample does include information on VSFs of material in the cloud.

\section{Results}\label{results}

In this section, we present our results on how VSFs reflect the velocity structure within and around MCs.

\subsection{Examples}\label{results:example}

\begin{figure*}[!htb]
	\centering
    \includegraphics[width=\textwidth]{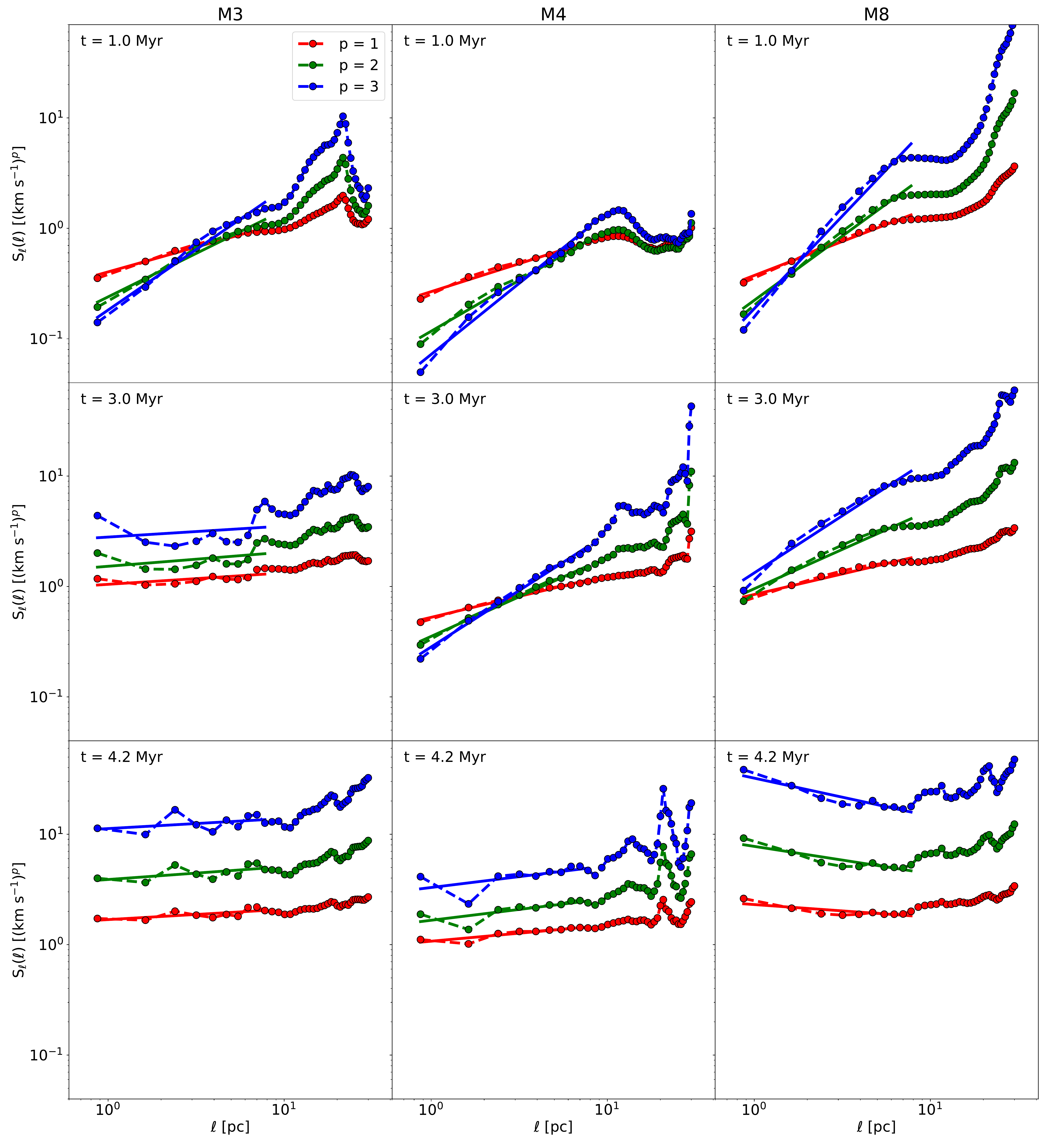}
    \caption{
        Examples of VSFs from models (\textit{left} to \textit{right}) \texttt{M3}, \texttt{M4}, and \texttt{M8} as function of lag scale $\ell$ and order $p$, based on data with density threshold. 
        The examples are given for times (\textit{top} to \textit{bottom}) t~=~1.0~Myr, 3.0~Myr, and 4.2~Myr.
        The dots (connected by dashed lines) show the values computed from the simulations. 
        The solid lines represent the power-law relation fitted to the respective structure functions.
    }
	\label{pic:results:vsf_example}
\end{figure*}

Fig.~\ref{pic:results:vsf_example} shows nine examples of density-weighted VSFs (Eq.~\ref{equ:method:def_vsf_dw}).
The figure shows the VSFs of all three clouds (columns) at times of 1.0, 3.0, and 4.2~Myr after the onset of gravity. 
All plots show orders $p=$1--3.
The solid lines show the fitted power-law relations as given by Eq.~(\ref{equ:method:fitting}).

The examples demonstrate that, in general, the measured VSFs cannot be described by a single power-law relation over the entire range of $\ell$.
Instead they are composed of roughly three different regimes: 
one at small scales at 0.8~pc~$\lesssim \ell \lesssim$~3~pc, a second one within 3~pc~$\lesssim \ell \lesssim$~10--15~pc, and the last one at large scales with 10--15~pc~$\lesssim \ell \lesssim$~30~pc.
We find that only the small and intermediate ranges may be represented by a common power-law relation.
On larger scales, one observes a local minimum before the VSFs either increase or stagnate.
Additional examples of VSFs are given in Appendix~\ref{appInertial}.

The examples in Fig.~\ref{pic:results:vsf_example} and Appendix~\ref{appInertial} illustrate how VSFs react to different scenarios that affect the turbulent structure of the entire clouds. 
All clouds at $t$~=~1.0~Myr show the case where turbulence is driven on large scales and naturally decays towards smaller scales.
This is the most common behaviour seen in all three MCs within the first $\sim$1.5~Myr of the simulations.
During this interval of time the clouds experience the effect of self-gravity for the first time in their evolution and need to adjust to this new condition.
Until this occurs, their VSFs continue to be dominated by the freely cascading turbulence that previously dominated the kinetic structure of the clouds.
We note that with each refinement it takes finite time for the turbulence to propagate to smaller scale, so the cloud evolution at high resolution was started well before self-gravity was turned on (see \citetalias{IbanezMejia2017}, \citealt{Seifried2017b}).

The later examples represent the clouds when the VSFs are dominated by sources that drive the flow within the clouds in a more coherent way.
\texttt{M3} at $t$~=~3.0~Myr and \texttt{M4} at $t$~=~4.2~Myr show VSFs at times when the clouds have just been hit by a SN blast wave.
One clearly sees how the amplitude of the VSFs are increased by an order of magnitude or more compared to the time before.
Especially the power at small scales below a few parsecs is highly amplified as a result of the shock.
Despite the increase of turbulent power at small scales, a large amount of energy is injected at large scales, as well.
However, the effect of SN shocks last for only a short period of time (see Sect.~\ref{results:normal}).

\texttt{M8} at $t$~=~3.0~Myr demonstrates the imprint of gravitational contraction.
Here, the VSF is almost flat, or even slightly increasing towards smaller separation scales. 
This kind of profile is typical for gas that is gravitationally contracting \citep{Boneberg2015,Burkhart2015}.
Gas moves into the inner regions of the cloud, reducing the average lag distances, while being accelerated by the infall to higher velocity.
As a consequence, large amounts of kinetic energy are transferred to smaller scales and higher densities, flattening the corresponding density-weighted VSF.

\subsection{Time Evolution}\label{results:normal}

\begin{figure*}[!htb]
	\centering  
  
  \begin{subfigure}[c]{\textwidth}
      \includegraphics[width=\textwidth]{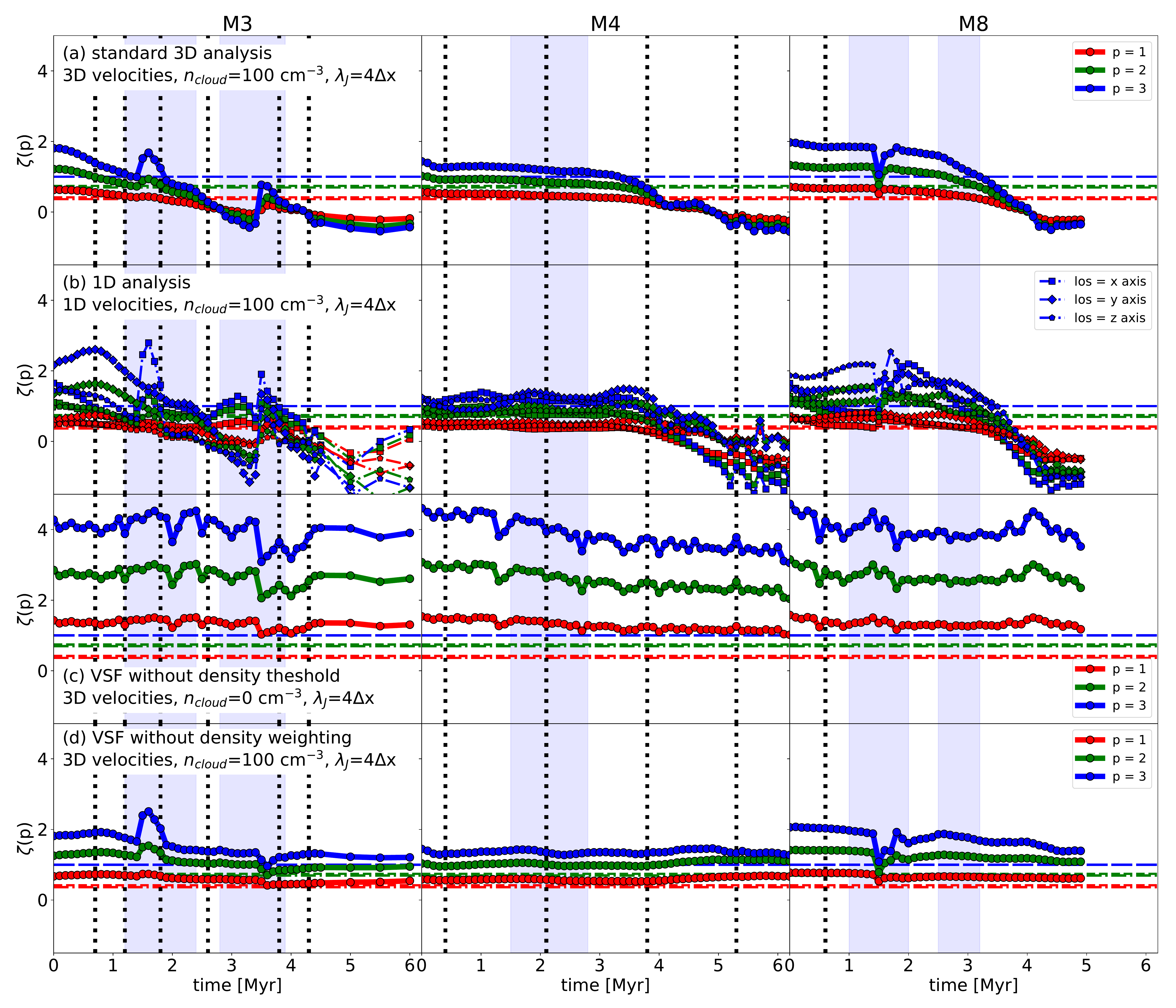}
      \label{pic:results:zeta_all_nojeans}
  \end{subfigure}
  \begin{subfigure}[c]{\textwidth}
      \addtocounter{subfigure}{4}
      \includegraphics[width=\textwidth]{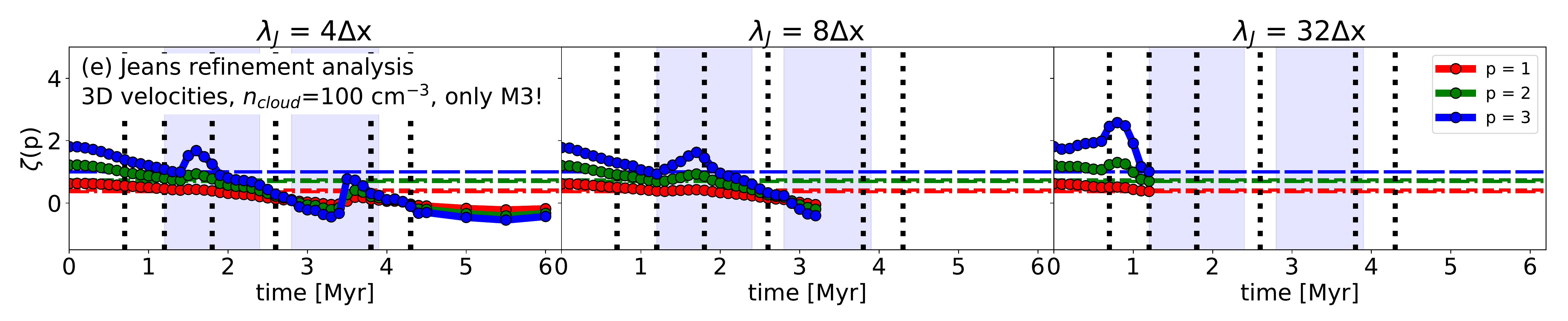}
      \label{pic:results:zeta_all_jeans}
  \end{subfigure}
  
  \caption{Time evolution of scaling exponent $\zeta(p)$ of the $p^\mathrm{th}$ order VSF. Panels (a)--(d) show the measurements for \texttt{M3}~(\textit{left}), \texttt{M4}~(\textit{middle}), and \texttt{M8}~(\textit{right}), respectively. Of these, (a) represents the standard analysis while the other rows illustrate the results of different variations in the analysis as noted in the figures. Panel (e) shows the values of $\zeta$ measured within \texttt{M3} as a function of the Jeans refinement level the cloud has been modelled with. Note that these more expensive runs were not run for as long as the fiducial run.  In all panels, the grey dotted vertical lines mark the times than a SN explodes within 100~pc of the cloud, while the blue areas indicate periods of enhanced mass accretion onto the clouds. The coloured horizontal lines show the predicted values for incompressible turbulence \citep[dash-dotted lines;][]{She1994} and for highly compressible, supersonic turbulence \citep[dashed lines;][]{Boldyrev2002}.
	\label{pic:results:zeta_all}
}
\end{figure*}

Fig.~\ref{pic:results:zeta_all}(a) summarises the time evolution of the power-law index $\zeta(p)$ fit to the density-weighted VSF obtained for each cloud, and each order $p$.
The figure shows several interesting features.
First, initially, at $t=0$~Myr, all calculated values of $\zeta$ are above the predicted values (see Eqs.~(\ref{equ:method:she}) and~(\ref{equ:method:boldyrev})).
This  probably occurs primarily because the base simulation only resolves down to 0.49~pc before additional grid levels are added to resolve the clouds, so it cannot resolve the turbulence inertial range below approximately 3 pc.
This can be seen in the $t=0$~Myr power spectrum in Fig.~25, Appendix B, of \citetalias{IbanezMejia2017}. 
As the zoom-in simulations evolve, the turbulence cascades to smaller scales in dense regions that are better resolved, so the density-weighted VSF slopes initially drop to the values expected for supersonic turbulence. 
The VSFs without density-weighting and, even more so, without density threshold remain too steep despite the increased resolution in dense regions, because these VSFs remain dominated by numerical diffusion in the diffuse gas that has not been further refined.

Second, $\zeta$ for all orders decreases with time as the clouds are first refined and then begin to gravitationally collapse. Distributed gravitational collapse causes an increase in relative velocities at increasingly small scales as material falls into filaments and nodes.  The increase in small-scale power leads to a flattening or even inversion of the VSF and thus a decrease in $\zeta$.
Third, occasionally one observes bumps and dips in slope in all orders of VSFs (e.g., \texttt{M3} or \texttt{M8} around $t=$1.7~Myr). 
These features only last for short periods of time (up to 0.6~Myr), but set in rather abruptly and represent sudden changes in large-scale power that change the VSF slope.

\begin{figure*}[!htb]
	\centering  
  
  \begin{subfigure}[c]{\textwidth}
      \includegraphics[width=\textwidth]{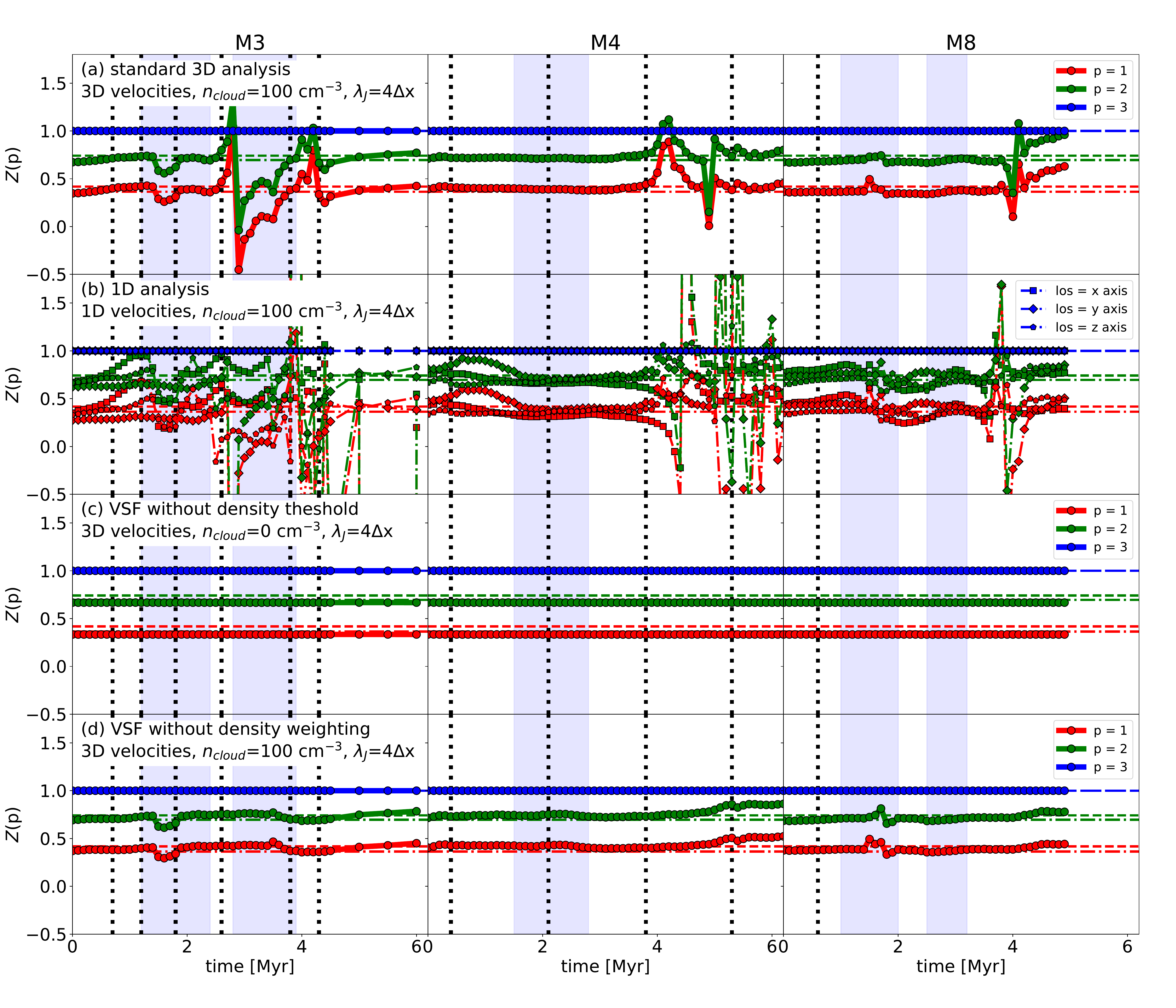}
      \label{pic:results:z_all_nojeans}
  \end{subfigure}
  
  \begin{subfigure}[c]{\textwidth}
      \addtocounter{subfigure}{4}
      \includegraphics[width=\textwidth]{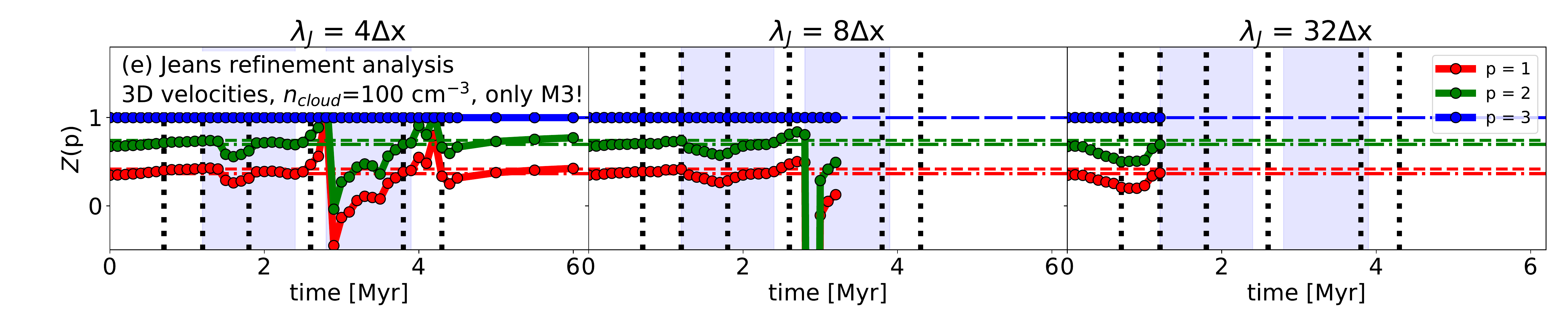}
      \label{pic:results:z_all_jeans}
  \end{subfigure}
  
  \caption{Like Fig.~\ref{pic:results:zeta_all}, but for the measured self-similarity parameter $Z = \zeta(p) / \zeta(3)$ of the $p^\mathrm{th}$ order VSF.}
	\label{pic:results:z_all}
\end{figure*}

Fig.~\ref{pic:results:z_all}(a) shows the corresponding time evolution of the self-similarity parameter, $Z$. 
One sees that most of the time the measured values of $Z$ are in agreement or at least closely approaching the predicted values. 
The occasional peaks in $Z$ (for example, in \texttt{M4} at $t=$4.1~Myr) occur at times when the scaling exponents of the VSFs $\zeta(3)$ reach values close to or below 0.
A decrease in $Z$ (for example, in \texttt{M3} around $t=$1.8~Myr) occurs when SN shocks hit and heavily impact the clouds, producing stronger effects in higher order VSFs.

Note that the nearby SNe that we mark in Figs.~\ref{pic:results:zeta_all} and~\ref{pic:results:z_all} and consider in the analysis were listed by \citetalias{IbanezMejia2017} as exploding at a radius of up to 100~pc from the clouds' centres of mass. The shock fronts move at average speeds of 50--100~km~s$^{-1}$ through the ISM, so it can take the blast waves more than 1~Myr to reach the clouds. Thus, it is important to keep in mind that the MCs do not react immediately to SNe, and that the time between the explosion of a SN and the interaction of its shock front with one of the clouds varies depending on the distance, as well as with the composition of the ISM along the propagation path.

\citetalias{IbanezMejia2017} discuss the influence of these SNe on the modelled clouds.
In particular, they focus on the overall properties of the clouds, like total mass, accretion rate, total velocity dispersion, and evolution.
They find that there is a tight correlation between those properties and that the clouds evolution strongly depends on whether the clouds are hit by blast waves, as well as the details of these interactions (meaning, for example, the strength or direction of the shock jump).
Therefore, it is intuitive that blast waves have a substantial influence on the turbulence within the clouds, as well. 
We will discuss this in more detail in the following sections of this paper.

In the rest of this section, we study how VSFs computed in different ways compare to these density weighted results.
We compare the findings with the results we have obtained with our original setup.
In Sect.~\ref{discussion}, we discuss and interpret these results in more detail.

\subsection{Line-of-Sight VSF}\label{results:1d}

Previously, we have seen how the VSF behaves and evolves within the clouds.
To do this, we derived the relative velocities based on the 3D velocity vectors from the simulations.
However, observed VSFs can only be measured using line-of-sight velocities.
In this subsection we investigate how VSFs derived from 1D relative velocities compare to the 3D VSFs presented before.

Figures~\ref{pic:results:zeta_all}(b) and~\ref{pic:results:z_all}(b) show measured $\zeta$ and $Z$, respectively, derived based on Eq.~(\ref{equ:method:def_vsf_1d}). 
We see that, in most of the cases, most of the 1D VSFs agree well with each other, as well as with the corresponding 3D VSFs.
The VSFs of \texttt{M4} are the best examples for this.

However, there are also cases in which the 1D VSF temporarily evolves completely differently than the 3D VSF.
For example, the 1D VSF along the x-axis in \texttt{M3} initially behaves like the corresponding 3D VSF, though with lower absolute values of $\zeta$ (or higher values of $Z$).
However, during the period $t=$2.5--3.8~Myr the functions diverge. 
While the 3D $\zeta$ decreases further and switches sign, the $\zeta$ based on the 1D VSF along the $x$-axis shows a local maximum before converging with the 3D $\zeta$ again. 
Another example for temporal divergence are the VSFs of \texttt{M8} projected onto the x- and z-axes. 
Here, the VSFs develop below or above the level of the 3D VSF, respectively.
After the impact of the SN blast wave at $t=$1.5~Myr all three VSFs converge with each other, as well as with the 3D VSF under the influence of the on-going gravitational collapse.

\subsection{Density Thresholds}\label{results:densthres}

We now examine the VSFs of the entire data cubes without setting a density threshold (i.e. setting $n_\mathrm{cloud}~=$~0~cm$^{-3}$).  Figure~\ref{pic:results:zeta_all}(c) shows $\zeta$, while Figure~\ref{pic:results:z_all}(c) shows $Z$ in this case.
These figures clearly illustrate that the measurements in the samples without density threshold completely differ from those with the density threshold.
The measured values of $\zeta$ are much higher in the ISM than in the cloud-only sample.
Although we see a slight decline of $\zeta$ in \texttt{M4} and \texttt{M8} as the gas contracts under the influence of gravity in the vicinity of the clouds, $\zeta$ generally evolves differently here than in the analysis that is focused on the clouds, remaining very steep.
We see a high rate of random fluctuations in the evolution of $\zeta$, as well.
Furthermore, contrary to all of our other test scenarios, we see here that all $Z$ are constant in time and within all clouds, with values slightly lower than those predicted by \citet{She1994} for incompressible flows.  
This is consistent with the high sound speed in the hot gas that fills most of the volume of the computational box, which results in subsonic flows predominating.

\subsection{Density Weighting}\label{results:densweight}

As mentioned in Sect.~\ref{methods:vsf}, Eq.~(\ref{equ:method:def_vsf}) represents the definition of the density-weighted VSF.
While this represents the observational situation better, the theoretical predictions were developed for the unweighted statistic.  Thus a comparison of results for the two variations in our model is of interest.
There are a few studies that have targeted this question 
\citep[e.g.,][]{Benzi1993,Schmidt2008, Benzi2010,Gotoh2002}.  
However, all of them considered isotropic, homeogeneous, turbulent flows that are not comparable to our clouds.
\citet{Padoan2016a} use both methods, but not on the same set of data. 

In this section, we investigate the influence of density weighting on VSFs by repeating the original analysis with the non-weighted VSF given by Eq.~\ref{equ:method:def_vsf}.
Figs.~\ref{pic:results:zeta_all}(d) and~\ref{pic:results:z_all}(d) show the measured values of $\zeta$ and $Z$ derived from the non-weighted VSFs, respectively.

Comparing the weighted and non-weighted samples, we see the following:
The non-weighted $\zeta$ (Fig.~\ref{pic:results:zeta_all}d) traces the interactions between the gas of the clouds and the SN shocks in the same way as occurs for the density-weighted VSF.
In \texttt{M3} and \texttt{M8} we also see that the values of $\zeta$ decrease as the clouds evolve, yet not as fast as they do in the density-weighted VSFs, which focus attention on the dense regions of strongest collapse. 
The measurements in \texttt{M4}, however, are almost constant over time. 
In all the cases, the values of $\zeta$ never decrease below 0.5; a behaviour that clearly differs from what we have observed in the density-weighted VSFs. 
The density weighting weakens the influence of the highly compressible flows in the densest regions, but not so much as in the case with no density threshold. 
Consequently, the evolution of $Z$s (see Fig.~\ref{pic:results:z_all}d) becomes smoother, as well, as there is no sign inversion of $\zeta$.
As a result the values of $Z$ fluctuate slightly when shocks hit, and otherwise vary between the compressible and incompressible limits.

\subsection{Jeans Length Refinement}\label{results:refinement}

The results we have discussed so far are based on simulation data presented in \citetalias{IbanezMejia2016} and \citetalias{IbanezMejia2017}.
Due to the huge computational expense of the variety of physical and numerical processes (fluid dynamics, adaptive mesh refinement, SNe, magnetic fields, radiative heating and cooling, and many more) within those simulations, though, they have required some compromises.
One of these compromises was the Jeans refinement criterion used as part of the adaptive mesh refinement mechanism.
The authors resolved local Jeans lengths by only four zones ($\lambda_J=4\Delta x$).
This is the minimal requirement for modelling self-gravitating gas in disks in order to avoid artificial fragmentation \citep{Truelove1998}. 
Other studies, have shown that a significantly higher refinement is needed to reliably resolve turbulent structures and flows within disks to resolve turbulence \citep{Federrath2011, Turk2012}.  
In our case, the key question is how quickly the turbulent cascade fills in after the multiple steps of refinement to higher resolution required to develop the high resolution cubes we use.  
Although we have a different physical situation, the earlier results still emphasise the importance of how well the Jeans length is resolved.

In the appendix of \citetalias{IbanezMejia2017}, the authors examine the effect the number of zones used for the Jeans refinement has on the measured kinetic energy.
For this, they have rerun the simulations of \texttt{M3} twice; 
once with $\lambda_J=8\Delta{}x$ for the first 3~Myr after self-gravity was activated, and once with $\lambda_J=32\Delta{}x$ for the first megayear of the cloud's evolution.
The authors show that the $\lambda_J=32\Delta{}x$ simulations smoothly recover the energy power spectrum on all scales already after this first megayear.
The other two setups do this, as well, but only over longer timescales (see also \citetalias{IbanezMejia2017}, \citealt{Seifried2017b}).

Furthermore, \citetalias{IbanezMejia2017} calculated the difference in the cloud's total kinetic energy as a function of time and refinement level.
They found that the $\lambda_J = 4\Delta{}x$ simulations miss a significant amount of kinetic energy, namely up to 13\% compared to $\lambda_J = 8\Delta{}x$ and 33\% compared to $\lambda_J = 32\Delta{}x$.
However, they also observed that these differences peak around $t=0.5$~Myr and decrease afterwards, as the $\lambda_J = 4\Delta{}x$ and $\lambda_J = 8\Delta{}x$ simulations adjust to the new refinement levels.
Thus, the results we have derived from the $\lambda_J = 4\Delta{}x$ simulations need to be evaluated with respect to this lack of turbulent energy, although the clouds' dynamics remains dominated by gravitational collapse.
It also means that the $\lambda_J = 4\Delta{}x$ data become more reliable the longer the simulations evolve.

\begin{figure*}
	\centering
    \includegraphics[width=\textwidth]{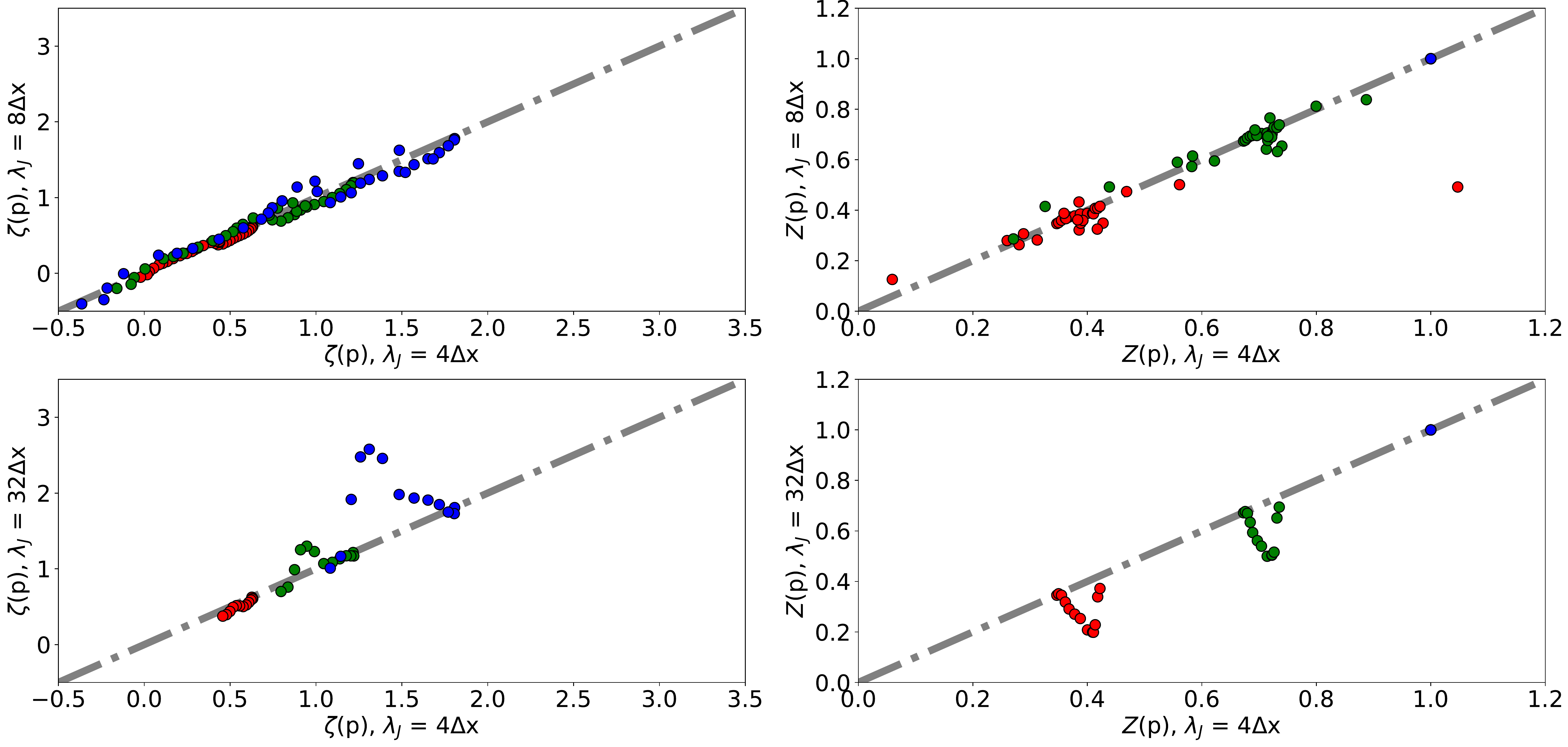}
    \caption{Comparison of VSF scaling exponents, $\zeta$ (\textit{left}), and self-similarity parameters, $Z$ (\textit{right}), depending on the Jeans refinement of the simulation runs the data are based on. The abscissas give values from $\lambda_J = 4\Delta{}x$, while the ordinates give values from $\lambda_J = 8\Delta{}x$ (\textit{top}) and $\lambda_J = 32\Delta{}x$ (\textit{bottom}). All data points refer to the \texttt{M3} cloud and represent different lags in the same time step in the respective simulations. 
    }
    \label{pic:results:jeans_comp}
\end{figure*}

In order to study  how the level of Jeans refinement influences the behaviour of the VSFs, we investigate the \texttt{M3} data of the $\lambda_J = 8\Delta{}x$ and $\lambda_J = 32\Delta{}x$ simulations.
Figs.~\ref{pic:results:zeta_all}e and \ref{pic:results:z_all}e show $\zeta$ and $Z$ for  $\lambda_J = 8\Delta{}x$ and $\lambda_J = 32\Delta{}x$.
In Fig.~\ref{pic:results:jeans_comp} we directly compare the measurements of all refinement levels relative to $\lambda_J = 4\Delta{}x$.

The $\lambda_J = 8\Delta{}x$ model shows the same behaviour as $\lambda_J = 4\Delta{}x$, with values in both samples being in good agreement as the top panel of Fig.~\ref{pic:results:jeans_comp} demonstrates. 
Over the entire observed time span, the measured values of $\zeta$ decrease as the VSF become flatter.
At the time the SNe interact with the cloud, over the course of about a megayear after traveling across the distance from the point of explosion to the cloud, the VSFs steeply increase toward larger scales, causing values of $\zeta$ (Fig.~\ref{pic:results:zeta_all}e) to jump.
Compared to the $\lambda_J = 4\Delta{}x$ sample, the peak in $\zeta$ is smoother and lasts longer at higher Jeans resolution.

These same effects can be seen in Fig.~\ref{pic:results:z_all}e where the drop of $Z$ due to the SN shock lasts longer than it did for $\lambda_J = 4\Delta{}x$. 
Besides this, the time evolution of $Z$ for $\lambda_J = 8\Delta{}x$ is as sensitive to the turbulence-related events as it was for $\lambda_J = 4\Delta{}x$.
The divergence produced when gravity has transferred the majority of power to smaller scales occurs at the same time. 
The actual depth of the drop is a numerical artefact caused by $\zeta(3)$ being equal or close to zero at this very time step. 

The picture changes when we analyse the VSFs based on the $\lambda_J = 32\Delta{}x$ runs (Figs.~\ref{pic:results:zeta_all}e,~\ref{pic:results:z_all}e, and~\ref{pic:results:jeans_comp}).
Here one sees that the measured values of both $\zeta$ (Fig.~\ref{pic:results:zeta_all}e) and $Z$ (Fig.~\ref{pic:results:z_all}e) are similar to those for $\lambda_J = 4\Delta{}x$ for the first 0.2~Myr.
After this short period, though, the evolution of $\zeta$ diverges. 
While $\zeta(1)$ and $\zeta(2)$ continue to decrease similar to $\lambda_J = 4\Delta{}x$ but at lower rate, $\zeta(3)$ increases until it peaks at $t=0.8$~Myr and falls steeply again.
This divergence has notable impact on the evolution of $Z$, as well. 
The bottom panel of Fig.~\ref{pic:results:jeans_comp} illustrates the different evolution of measured $\zeta$ and $Z$ in the two simulations more clearly.
One sees that the differences between the samples follow the same pattern for all orders of $p$.
The differences, though, increase with the order:
While the values for $\zeta(1)$ are still in good agreement, the measured values of $\zeta(2)$ and $\zeta(3)$ for $\lambda_J = 32\Delta{}x$ are 40\% and 100\% higher than those measured for $\lambda_J = 4\Delta{}x$, respectively.
Consequently, this causes differences in $Z(p)$ of 30--52\% between the simulations.
At $t=$1.2~Myr, the last time step of this sample, the values of all $\zeta$ equal the measurements of $\lambda_J = 4\Delta{}x$ again.
As the cost of extending the $\lambda_J = 32\Delta{}x$ simulation is prohibitive, we cannot determine whether this agreement will continue.

\section{Discussion}\label{discussion}

\subsection{Time evolution}\label{discussion:normal}

We have seen in Sect.~\ref{results} that density-weighted VSFs reflect a combination of uniform, compressible turbulence, large-scale shocks, and gravitational collapse.  Extended self-similarity emphasises the turbulent nature of these high-Reynolds numbers flows even in regions of gravitational collapse. 
The measured values of $\zeta$ differ from the predicted values by \citet{She1994} and \citet{Boldyrev2002} for most of the time of the clouds' evolution.
The origin of these differences are to be found in a combination of numerical dissipation and the underlying physical conditions. 
Contrary to the conditions in the reference models, we examine a multi-phase medium with magnetised turbulence that is hardly uniform, isothermal, or homogeneous.
However, the diffuse medium is also underresolved compared to the dense gas, thanks to our Jeans refinement strategy, which removes small-scale power in the diffuse medium.

The impact of SN shocks hitting the clouds is to inject power at all scales (Fig.~\ref{pic:results:vsf_example}). 
The resulting VSFs tend to lose their power-law character. Fitting a power-law to them anyway results in substantial perturbations from the predictions for compressible turbulence even under extended self-similarity.
Fig.~\ref{pic:results:z_all} shows times of SN explosions and periods of strong accretion onto the clouds. 
Remembering that it can take the shock front more than 1~Myr to propagate from the site of the SN explosion to the molecular cloud, perturbations in $Z$ not associated with zero-crossings by $\zeta(3)$ are consistent with being caused by SN shock front interactions with the clouds.  
These shock interactions last for only a fraction of a megayear, though, consistent with the crossing time of the blast wave through the dense interior of the cloud, after which the turbulent nature of the flow reasserts itself.

As the clouds gravitationally collapse, the resulting increase in small-scale power flattens or even inverts the density-weighted VSFs, resulting in decreasing or even negative values of $\zeta$ (Fig.~\ref{pic:results:zeta_all}(a)). The increase in small-scale power can also be derived from the increasingly negative binding energy of the clouds as further gas falls into them \citepalias{IbanezMejia2017}. 
At the same time as the turbulence becomes increasingly non-uniform and anisotropic because of the importance of gravitation, the bulk velocity dispersion of the cloud increases.
\citetalias{IbanezMejia2016} showed that Eq.~(\ref{eq:larson}) is satisfied at these late times, but not at early times, less than a free-fall time, when the velocity dispersion inherited from the background turbulent flow is independent of the size of the cloud. 
These early times are when the turbulence dominates the flow and the second-order power law is roughly $\zeta(2) \simeq 1/2$.
This suggests that the apparent agreement with Larson's size-velocity relationship is coincidental. Observing two-point correlations using a method that cannot capture the dense flows adequately will yield this result from the cloud envelopes, though, thus perhaps explaining the apparent success of such efforts \citep[][Fig.\ 9 shows how these different interpretations can arise]{Goodman1998}.

Extended self-similarity shows VSF ratios characteristic of compressible turbulence (Fig.~\ref{pic:results:z_all}a), as can be seen from their tending to lie between the incompressible limit of \citet{She1994} and the extremely compressible Burgers turbulence limit of \citet{Boldyrev2002}.
(The extended self-similarity procedure fails as $\zeta(3)$ passes through zero, however, so it must be interpreted in concert with the raw values of $\zeta$.)
This suggests that, just as the extended self-similarity procedure removes the effects of dissipation, it also removes the effects of hierarchical gravitational collapse, while continuing to reflect the turbulence in these high Reynolds-numbers flows.

\subsection{Line-of-sight velocities}\label{discussion:1d}

In Sect.~\ref{results:1d} we have seen that the $\zeta$ and $Z$ derived from the 1D VSFs generally evolve similarly to those derived from the 3D VSFs.
Yet, we have also seen that individual sight lines may evolve differently.
These differences appear to reflect the detailed geometry of shock impacts on the cloud, which are reflected more strongly in the higher-order VSFs.
For example, for the first 2~Myr of the evolution of \texttt{M4} the values of $Z$ along the $y$-axis are significantly higher than those observed along the other axes and diverge significantly from the values expected for uniform turbulence.
Recall that a perturbation in $Z$ usually corresponds to an episode of strong shock driving, suggesting an impact along the $y$-axis at this time. 
Along the other two axes, $Z$ continues to agree with supersonic turbulence \citep{Boldyrev2002}.
This effect is only visible as we analyse the three dimensions separately, while the driving of the gas along the $y$-axis is averaged out in the 3D VSFs (see Fig.~\ref{pic:results:z_all}a).

In summary, for a fully developed 3D turbulent field we expect that 1D VSFs behave similarly to 3D VSFs.
However, when there is a preferred direction along which the gas flows, the 1D and 3D VSFs differ significantly from each other. 
Thus, we predict that observed VSFs reflect the nature of turbulence within MCs unless there is clear evidence that the gas is driven in a particular direction (e.g., by a stellar wind or SN shock front).

Note that this analysis does not take typical line-of-sight effects, such as optical depth or blending, into account. 
Future studies need to investigate this point in more detail by performing VSF analyses based on full radiative transfer calculations.

\subsection{Density thresholds}\label{discussion:densthres}

We find that the structure and behaviour of VSFs strongly depends on whether or not we assume a density threshold in computing them.
In the fiducial case, where n$_\mathrm{cloud}$~=~100~cm$^{-3}$, we have seen a mostly straight decline of $\zeta$ while $Z$ remains fairly constant over time, reflecting the contraction of the clouds due to self-gravity.
On the other hand, if we remove the density threshold, including the entire high-resolution cube in the calculation, we observe a completely different picture.
The high velocities present in the diffuse interstellar medium surrounding the cloud lead to strong large-scale power and thus much steeper VSFs, corresponding to higher values of $\zeta$. 
There is still a slightly declining trend in $\zeta$, but the evolution is dominated by random fluctuations.
We also see that $Z$ remains constant for the entire simulation in every case.
However, the VSF scaling exponents are about four times steeper than the values that are predicted by \citep{Boldyrev2002} for incompressible flows.
This is partly an effect of numerical dissipation reducing the velocities in less-refined diffuse gas.
However, it is probably also due to the sharp reduction in the velocity dispersion of dense gas already noted in \citetalias{IbanezMejia2016}, given that the dense gas has small characteristic scales.
This suggests that they are dominated by the subsonic flow in the hot gas with $T > 10^6$~K that occupies most of the volume of the box.

Furthermore, the results demonstrate that the effect of SNe, and the interaction of the produced shock fronts with the ISM acts rather locally. 
This means that a single SN can not drive the gas dynamics on scales of our entire cubes (40$^3$~pc$^3$), at least not in the same way as it does on scales of individual MCs.
Rather the VSFs reflect the superposition of multiple SNe that only combined drive the turbulence of the diffuse ISM.

We conclude that the decision of whether or not a density threshold is used has a significant and direct influence on the resulting VSFs.
Indeed, it is a straight-forward approach to focus the analysis on the actual area of interest.
In observational studies such a threshold will anyway always be present as minimal collision rates for excitation or the sensitivity of detectors automatically introduce implicit density or intensity thresholds. 
Although we have only tested two specific setups in this context we have seen the significance of a proper choice of the density threshold, as well as a proper discussion of the obtained results considering the threshold as one of the defining parameters.
We note that future work could fruitfully also consider an upper density threshold to mimic the major effects of optical thickness even if true radiative transfer were probibitively expensive.

\subsection{Density weighting}\label{discussion:densweight}

In this subsection, we discuss the effect of computing the VSF with or without including density weighting, relying on the results presented in Sect.~\ref{results:densweight}.
As long as the turbulence is dominated by the large scales, and a density threshold is used, considering the density weighting does not have a significant effect.
However, as the clouds evolve the differences increase:
the non-weighted VSFs never drop below 0.5.
This is because the non-weighted VSF treats all cells equally, no matter whether the particular cell represents a dense element of the cloud centre or a diffuse element of the cloud's edge, while the weighted VSF gives more weight to the matter within the small-scale, dense, collapsing regions.
The kinetic energy is dominated by these regions.
Thus, neglecting density weighting decouples the VSF from the kinetic energy distribution.
(A more exact treatment of this question is given by \citet{Kritsuk2013a} and \citet{Banerjee2017,Banerjee2018}.)
This is particularly important at late times when small-scale collapse dominates.

\begin{figure*}
	\centering
    \includegraphics[width=\textwidth]{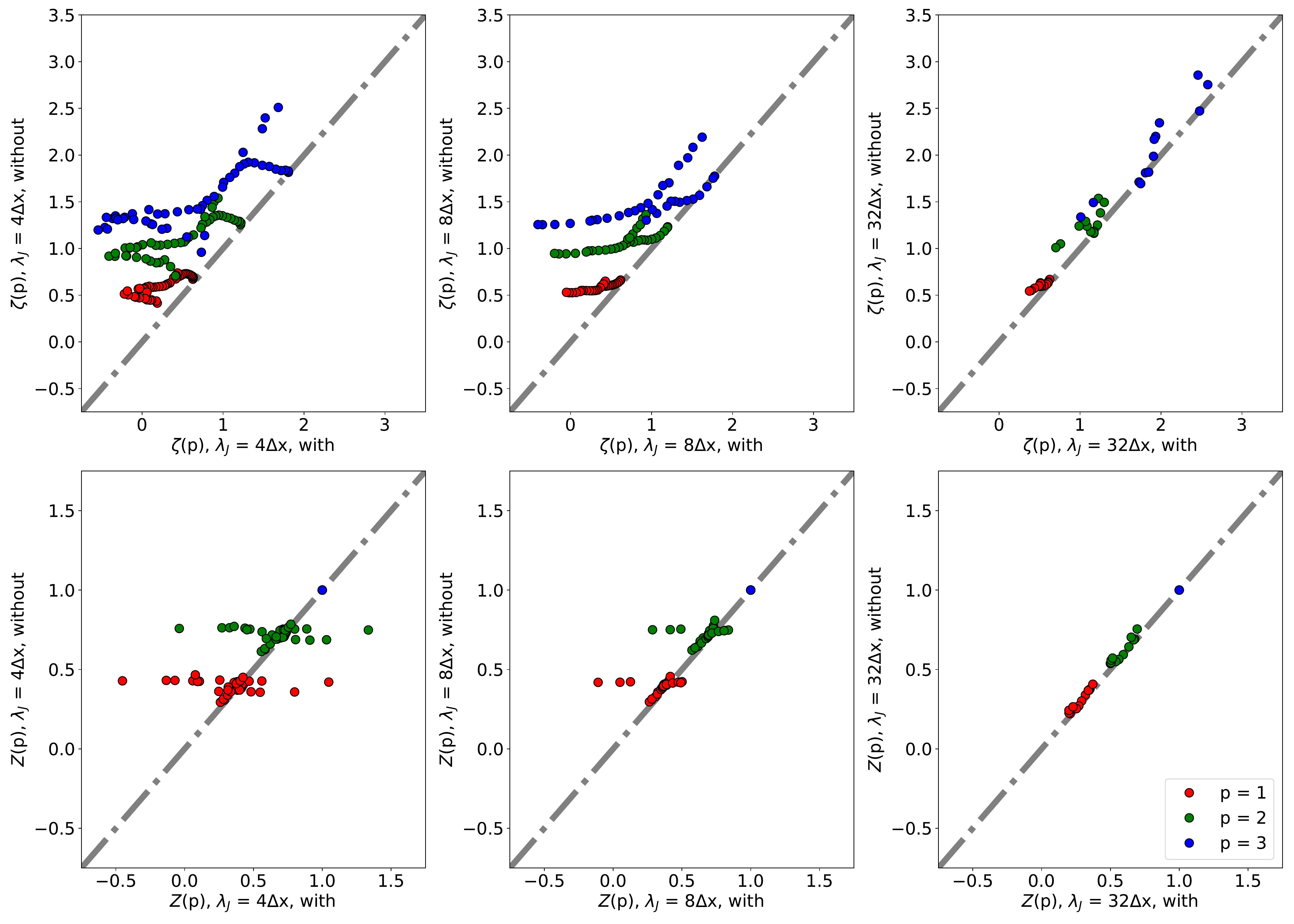}
    \caption{ Comparison of $\zeta$ (\textit{top}) and $Z$ (\textit{bottom}) measured based on density-weighted VSFs (\textit{abscissas}) and non-weighted VSFs (\textit{ordinates}). We note that the given values are based on \texttt{M3} only.}
    \label{pic:results:comp_weighting}
\end{figure*}

Nevertheless, Fig.~\ref{pic:results:z_all}(d) illustrates that these differences do not prevent extended self-similarity from holding. 
Regardless of whether density-weighting is included, the values of $Z$ remain similar, with similar responses to external driving, except for the features created when $\zeta(3)$ passes through zero in the density-weighted VSFs.
This observation is true for all Jeans refinement levels, as Fig.~\ref{pic:results:comp_weighting} demonstrates.

Fig.~\ref{pic:results:comp_weighting} summarises the comparison of $\zeta$ and $Z$ measured with the density-weighted and non-weighted VSFs for all Jeans refinement levels (meaning the granularity used for modelling the turbulent motions of the gas, see Sect.~\ref{results:refinement} for more details).
The figure clearly shows that the measurements only agree well for the highest refinement level with $\lambda~=~32\Delta x$.
However, we would need more data points to be sure that this correlation is indeed real.
At lower refinement levels the measurements, as those used for the standard analysis and all other test scenarios but the one presented in Sect.~\ref{results:refinement}, correlate less well with each other. 
The differences in the samples appear dominantly when the density-weighted $\zeta$ cease below $\approx$0.5, which is the global minimum for the non-weighted $\zeta$. 
This means that none of the $\zeta$ computed in all clouds and refinement levels with the non-weighted VSF is measured to be below 0.5.

We conclude that deriving the VSF from smooth density distributions without considering density-weighting does not affect the behaviour of $\zeta$ and $Z$, as long as the turbulence is dominated by large scale flows, but it has a significant effect on the measurements when the small scales become dominant.
The latter is particularly important as this finding has a directly impact on the conclusions drawn based on the scales and mechanisms that drive the turbulence based on the measured $\zeta$.
Not only does $\zeta$ become insensitive to the influence of gravitational contraction with time, the non-weighted VSFs also does not reflect when the majority of kinetic energy has been transferred to small scales. 
Furthermore, this emphasises the importance of taking optical depth effects into account, as single tracers covering limited density ranges may effectively provide statistics closer to the non-weighted VSFs.

\subsection{Jeans length refinement}\label{discussion:refinement}

In Fig.~\ref{pic:results:jeans_comp} we see that the choice of refinement level has no significant influence on the measurements and evolution of both $\zeta$ and $Z$. 
The $\lambda_J=4\Delta{}x$ and $\lambda_J=8\Delta{}x$ models are in good agreement with each other.
This means that, although refining Jeans lengths with 4~cells misses about 13\% of kinetic energy, the effect on the structure and behaviour of the turbulence is rather small and not traced by the VSF analysis.

However, Fig.~\ref{pic:results:jeans_comp} shows that the agreement is rather poorer with $\lambda_J=32\Delta{}x$, as the latter differs more from $\lambda_J=4\Delta{}x$ the higher the order of the VSF is.
Following the explanations in Sect.~\ref{results:refinement}, the behaviour of $\zeta$ and $Z$ in the $\lambda_J=32\Delta{}x$ runs corresponds to the reaction of the cloud's gas to a shock wave running through the cloud; caused by a SN that exploded before $t$~=~0~Myr. 
Indeed one sees a SN at a distance of 172~pc at $t=-1.11$~Myr. 
As the power of the shock decreases rapidly with distance, the SN is too weak to effectively compress the gas within \texttt{M3}.
This is why it was not detected in the less refined samples.

The SN explodes far below the mid-plane of the simulated disk galaxy, in a region without dense gas, so the blast wave remains strong as it propagates through the ISM. 
By the time the blast arrives at cloud \texttt{M3}, it is still energetic enough to impact the cloud with winds at velocities above 300~km~s$^{-1}$, at the closer edge of the cloud. 
This causes an increase of VSFs at longer lag scales and the increase of $\zeta$, as well as the drop in $Z$.
However, in the less refined runs the arrival of the blast wave only contributes to the normal external driving for the gas' turbulence.
Only the $\lambda_J=32\Delta{}x$ runs can capture the fine structure of the shock driven into the cloud by the blast wave, which is why we detect this feature in these runs only. 
This emphasises the importance for studies like ours of properly resolving the full internal structure of realistic MCs.

We conclude that improving the resolution resolves details that can affect the VSF, but that the overall behaviour is already determined by our moderate resolution simulations.

\subsection{Comparison to observations}\label{discussion:observation}

The majority of studies of VSFs in MCs are based on simulated data, as is the work presented in this paper.
However, there are also some surveys that derive VSFs from observations, or whose data can be used to reconstruct the scaling properties of VSFs. 
In this section we discuss our results in the context of the observational studies that we list in Table~\ref{tab:discussion:summary_obs}.

\begin{table*} 
\centering 
	\begin{tabular}{llccc} 
	\centering 
		Reference 	& Comments & p & $\zeta$ & Z \\ \hline  \hline
        \citet{Heyer2007}	& $^{12}$CO J = 1-0, & 1 &  0.49 $\pm$  0.15 &  --- \\ 
                            & Perseus \& Solar Neighborhood & & & \\ \hline
        \citet{Heyer2015} 	& $^{12}$CO \& $^{13}$CO J = 1-0, 30  MCs & 1 &  0.24 $\pm$ 0.00 & --- \\ 
                            & $^{12}$CO \& $^{13}$CO J = 1-0, Taurus &  1 &  0.26 $\pm$  0.00 & --- \\  \hline
        \citet{Miesch1994} 	& $^{13}$CO J = 1-0, & 1 &  0.43 $\pm$  0.15 & --- \\ 
                            & 12 clouds and subregions of GMCs & 2 &  0.86 $\pm$  0.30 &  --- \\  \hline
        \citet{Padoan2003}	& $^{13}$CO J = 1-0, Perseus & 1 &  0.50 $\pm$  0.00 &  0.42 $\pm$  0.00 \\ 
                            &  & 2 &  0.83 $\pm$  0.00 &  0.72 $\pm$  0.00 \\ 
                            &		 & 3 &  1.18 $\pm$  0.00 &  1.00 $\pm$  0.00 \\ 
                            & $^{13}$CO J = 1-0, Taurus & 1 &  0.46 $\pm$  0.00 &  0.42 $\pm$  0.00 \\ 
                            &		 & 2 &  0.77 $\pm$  0.00 &  0.72 $\pm$  0.00 \\ 
                            &		 & 3 &  1.10 $\pm$  0.00 &  1.00 $\pm$  0.00 \\  \hline
        \citet{Padoan2006} 	& $^{13}$CO J = 1-0, Perseus & 2 &   0.80 $\pm$  0.10    &  --- \\  \hline
        \citet{RomanDuval2011} & $^{13}$CO J = 1-0, 367 clouds  from the GRS survey & 1 & 0.50  $\pm$   0.30  &   --- \\  \hline
        \citet{Zernickel2015} & $^{13}$CO J = 1-0, NGC 6334 & 1 &  0.38 $\pm$  0.00  &   --- \\ 
                            &		 & 2 &  0.76  $\pm$    0.01 &  --- \\ 
                            & $^{13}$CO J = 1-0, NGC 6334, $\ell \leq$ 4 pc  & 1 &  0.48 $\pm$  0.01 &  --- \\
                            & & 2 &  0.79 $\pm$  0.01 &  --- 
	\end{tabular} 
	\caption{Summary of observed $\zeta$ and Z in the literature.} 
	\label{tab:discussion:summary_obs} 
\end{table*}

Most of the velocity information derives from $^{12}$CO and $^{13}$CO observations of young star-forming regions \citep[e.g., Perseus and Taurus][]{Padoan2003}.
We also consider observations of more evolved regions, such as those of the H~{\sc ii} region NGC 6334 \citep{Zernickel2015}, since the filaments in our simulated MCs fragment within the first 2 Myr \citepalias{Chira2018}, suggesting that our cloud may start to resemble such more evolved regions.

\begin{figure*}
	\includegraphics[width=\textwidth]{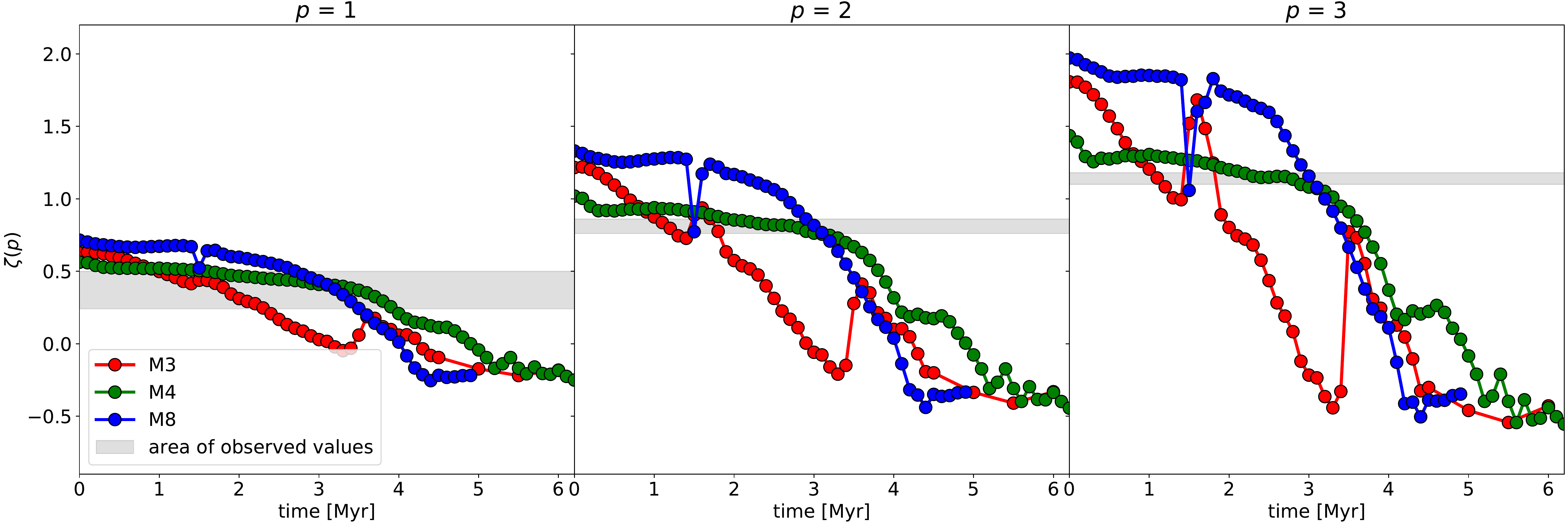}
	\caption{
	Time evolution of scaling exponent $\zeta(p)$ of the $p^{\rm th}$ order VSF. 
	The panels show the measured scaling exponents for the first to third order (\textit{left} to \textit{right}) for \texttt{M3} (red), \texttt{M4} (green), \texttt{M8} (blue). 
	The shaded area within all panels marks the range of observed values at the respective order of VSF (see Table \ref{tab:discussion:summary_obs}).
	}
	\label{pic:discussion:comp_observation}
\end{figure*}

Fig.~\ref{pic:discussion:comp_observation} summarises the measured scaling exponents found in the literature, along with our fiducial results (Fig.~\ref{pic:results:zeta_all}a).
We see that the observed values of $\zeta$ span within values of 0.24 to 1.18 across the first three orders. 
Furthermore, we see that the observed values are always positive, suggesting that none of the observed clouds show signs of being dominated by collapse, though that could be because the fastest flows lie in optically thick regions inaccessible to the observations.  

Compared to these measurements, the distributions of our results show a large scatter across the parameter space. 
However, we also see that there is a significant fraction of values in our models that are consistent with the observational findings. 
These measurements belong roughly to the evolutionary stages of the modelled clouds after having evolved for roughly 1.0--3.9~Myr after the onset of self-gravity.
\citetalias{Chira2018} finds that the clouds consist of a highly hierarchical structure that is dominated by already fragmenting filaments at this point.
This means that the flows within the clouds experience a transition from cloud-scale dominated, through filament-dominated, to core-collapse driven motions; which is exactly what we observe in the VSFs, as well.
Consequently this suggests that the observed MCs, which show clear signs of embedded star formation activity, are in a similar stage where flows are dominated by the formation of hierarchical structures, (pre-/proto-)stellar cores or, in the case of NGC 6334, internal feedback.

We find that the interpretation of observational measurements is still difficult for several reasons:
\begin{enumerate}
\item We have already discussed in Sect.~\ref{discussion:1d} that the transformation from 3D to 1D VSFs is not trivial, in particular when the studied flows are not isotropic.
This is, for example, the case when the first structures (such as filaments or sheets) form, or the first cores collapse and accrete.
\item We have seen that interactions with SN shocks may trigger a preferred direction, that has the potential to strongly influence the measured 1D VSF.
Although the influence of shock fronts on VSFs is transient compared to the lifetime of the entire MC, it is still long enough to mimic a quasi-steady state in real observations.
Observing typical shock tracers, such as SiO, may help to identify these situations. 
\item We have neglected typical line-of-sight effects that may have a significant influence on the measurements of the local standard of rest velocity whose precision is crucial for this kind of study.
Our projections ignore optical depth effects, and reflect velocities all the way through the clouds, including high column density regions of dynamical collapse where motions are fast at small scales.
However,  both $^{12}$CO and $^{13}$CO reach optical depth of unity at relatively low column densities. 
This means that the observed VSFs will only reflect the motions of the surface layers of dense MCs. 
Fig.~\ref{pic:results:zeta_all}, as well as the figures in Appendix \ref{appFitting} demonstrate that neglecting certain regimes, in particular at small scales where column densities are highest, does have significant influence on the shape and scaling of the corresponding VSF.
Therefore, single-tracer observations are not suitable for studying the dynamical structure of MCs. 
For a proper VSF analysis it would be advisable to use a variety of tracers to cover the different phases of the clouds, as well as to populate the statistics of lag distances more completely.
\item Only a small fraction of the listed observational studies in Table~\ref{tab:discussion:summary_obs} aimed to measure the VSFs of the respective objects directly.
In the majority of cases, the focus of the investigations was on the general budget of kinetic energy within the MC, as well as the question whether those clouds follow Larson's size-velocity relation (Eq.~[\ref{eq:larson}]).
It is unclear whether the difference between a relation of the lag distance of two particles and their relative line-of-sight velocity and the connection between the size of the entire MC and the velocity dispersion of the contained gas has always been considered.
\end{enumerate}

We recommend that both theorists and observers discuss in more detail how observational studies may use VSFs in the future.
From the theoretical point-of-view, full line radiative transfer calculations are required to better evaluate observational biases and simple projection effects.
This requires observations with a high spatial resolution of the respective MC for a wide range of lag scales and good statistics for fitting the scaling of VSF, as well as lines with well-defined line-of-sight velocities, ideally, optically thin lines of intermediate- and high-density tracers.

\section{Summary \& Conclusions}\label{conclusions}

In this paper, we analyse the VSFs of MCs that have formed within 3D magnetohydrodynamical, adaptive mesh refinement, FLASH simulations of the self-gravitating, SN-driven ISM by \citetalias{IbanezMejia2017}, including both density weighting and a density cutoff.
The main results are as follows.

\begin{enumerate}
	\item The scaling of the density-weighted VSFs depends on a combination of turbulence and more coherent processes such as SN blast wave impacts and gravitational contraction. We find that the power-law scaling $\zeta$ of 3D density-weighted VSFs reflects the development of gravitational contraction, while the extended self-similarity scaling $Z$ reveals interactions of clouds with large-scale blast waves.
    \item The two different proposed explanations for Larson's size-velocity relationship, a turbulent cascade and gravitational contraction,  appear to apply to different stages in the evolution of MCs, as well as different observational techniques. It appears {\em coincidental} that they have the same functional relationship of length to velocity, which has led to confusion of one with the other.
    \begin{itemize}
        \item MCs dominated by uniform turbulence show a first-order VSF with $\zeta(1) \simeq 1/2$.
        The same result can be found for clouds undergoing strong gravitational contraction by computing the VSF without density weighting (or, most likely, in the presence of optical depth effects), which is dominated by the low-density, turbulence-dominated outer regions of clouds.
        At the initial time, though, we measure values of $\zeta(2)$~=~1 or larger in the density-weighted VSFs, that are inconsistent with turbulence models or simulations that predict $\zeta(2) = 0.74$ (Eq.~\ref{equ:method:boldyrev}).
        However, these initial values are affected by numerical dissipation and tend to decrease as the resolution in dense regions increases in the zoom-in runs.
        \item Examining the overall velocity dispersion of gravitationally dominated clouds undergoing star formation, on the other hand, reflects the dynamics of gravitational collapse.
        In this case, the cloud shows a shallow or even inverted VSF dependence $\zeta(1) \lesssim 0$. 
        This reflects strong flows at small scales. However, such gravitationally contracting clouds were shown by \citetalias{IbanezMejia2016} to have an overall square-root velocity-radius relationship (Eq.~[\ref{eq:larson}]) given by free-fall or virial equilibrium (which differ by only $2^{1/2}$, as noted by \citealt{Ballesteros2011}).
    \end{itemize}
	\item As long as the MC is not affected by a shock, $Z$ agrees well with predicted values for supersonic flows, even as gravitational collapse proceeds.
	\item We test the influence of Jeans refinement on the VSFs. We find that the absolute amount of kinetic energy does not influence the evolution of $\zeta$ and $Z$, but that better resolution of external shocks can produce changes in both quantities.
	\item Comparison of 3D to 1D VSFs shows differences in detail, but qualitative agreement in the behaviour of both $\zeta$ and $Z$, in particular when gravity dominates gas dynamics. 
	Thus, observed 1D VSFs can be useful diagnostics in gravitationally bound and contracting regions. 
	On the other hand, differences arise when strong transverse flows or shocks dominate the velocity field. 
	\item We calculate cloud VSFs using a density threshold to isolate the cloud material, as would characteristically happen in an observation of molecular material. 
	Without such a threshold, our VSFs are dominated by the diffuse ISM. In that case, the extended self-similarity scaling $Z$ lies just below the value predicted for isotropic, incompressible turbulence by \citet{She1994}. 
	This is consistent with the low Mach number in the hot, diffuse, ISM filling most of the volume of our simulation. \linebreak
    Yet, the actual values of $\zeta$ in the low density gas are about four times higher than those predicted by both \citet{She1994} and \citet{Boldyrev2002} due to some combination of numerical dissipation and the multiphase nature of the medium, which reduces velocities at small scale in dense regions (see Sect.~\ref{discussion:densthres}).
	\item We investigate the influence of defining the VSF with and without density weighting. 
	We find that the qualitative behaviour is traced by both approaches. 
	However, the scaling of the non-weighted VSF $\zeta$ is always positive, not falling nearly as far as for the density-weighted VSF. 
	The density-weighted VSF reflects the kinetic energy distribution better as gravitational collapse proceeds to smaller and smaller scales. 
	(Note that in, for example, CO observations, optical depth effects may obscure this behaviour.) 
	\item We compare our results with measurements of both $\zeta$ and $Z$ in observational studies. 
	We see that our findings are generally consistent with with observations within periods during which the clouds' flows are influenced by both turbulent flows and global gravitational contraction, including strong structure formation and starting fragmentation. This reflects the conditions of embedded star formation activity within observed MCs.
\end{enumerate}

Our analysis shows that VSFs are useful tools for examining the driving source of turbulence within MCs.
However, studies that use VSFs need to precisely review the assumptions and parameters included in their analysis as these can have a significant influence on the results.

For our simulated clouds, the VSFs illustrate that gravitational contraction dominates the evolution of the clouds.
During contraction, the VSF scaling exponent $\zeta(p)$ drops in value and can even become negative as kinetic energy concentrates on small scales.
Nevertheless, the extended self-similarity scaling parameters $Z(p)$ continue to agree with the analytic prediction for compressible turbulence except for short periods during which SN blast waves increase power on multiple scales.
Because such blast waves are neither homogeneous nor isotropic, they often lead to transient non-power law scaling of the VSFs, and thus strong departures from uniform turbulent behaviour of $Z(p)$.

\begin{acknowledgements} 
    We thank the anonymous referee for a detailed report that led to an improved and better focused exposition, particularly in the discussion of Larson's relation, as well as the inclusion of the Appendices.
    M-MML received support from US NSF grants AST11-09395 and AST18-15461, and thanks the A. von Humboldt-Stiftung for support.  
    JCI-M was additionally supported by the Deutsche Forschungsgemeinschaft (DFG) via the  Collaborative  Research  Center  SFB  956  ``Conditions and  Impact  of  Star  Formation'' (subproject  C5) and the  DFG  Priority  Program 1573 ‘The physics of the interstellar medium’.
\end{acknowledgements}

\bibliographystyle{aa} 
\bibliography{ref}

\begin{thebibliography}{58}
\expandafter\ifx\csname natexlab\endcsname\relax\def\natexlab#1{#1}\fi

\bibitem[{{Ballesteros-Paredes}
  {et~al.}(2011{\natexlab{a}}){Ballesteros-Paredes}, {Hartmann},
  {V{\'a}zquez-Semadeni}, {Heitsch}, \& {Zamora-Avil{\'e}s}}]{Ballesteros2011a}
{Ballesteros-Paredes}, J., {Hartmann}, L.~W., {V{\'a}zquez-Semadeni}, E.,
  {Heitsch}, F., \& {Zamora-Avil{\'e}s}, M.~A. 2011{\natexlab{a}}, \mnras, 411,
  65

\bibitem[{{Ballesteros-Paredes}
  {et~al.}(2011{\natexlab{b}}){Ballesteros-Paredes}, {V{\'a}zquez-Semadeni},
  {Gazol}, {Hartmann}, {Heitsch}, \& {Col{\'{\i}}n}}]{Ballesteros2011b}
{Ballesteros-Paredes}, J., {V{\'a}zquez-Semadeni}, E., {Gazol}, A., {et~al.}
  2011{\natexlab{b}}, \mnras, 416, 1436

\bibitem[{{Ballesteros-Paredes}
  {et~al.}(2011{\natexlab{c}}){Ballesteros-Paredes}, {V{\'a}zquez-Semadeni},
  {Gazol}, {Hartmann}, {Heitsch}, \& {Col{\'{\i}}n}}]{Ballesteros2011}
{Ballesteros-Paredes}, J., {V{\'a}zquez-Semadeni}, E., {Gazol}, A., {et~al.}
  2011{\natexlab{c}}, \mnras, 416, 1436

\bibitem[{{Banerjee} \& {Galtier}(2013)}]{Banerjee2013}
{Banerjee}, S. \& {Galtier}, S. 2013, \pre, 87, 013019

\bibitem[{{Banerjee} \& {Kritsuk}(2017)}]{Banerjee2017}
{Banerjee}, S. \& {Kritsuk}, A.~G. 2017, \pre, 96, 053116

\bibitem[{{Banerjee} \& {Kritsuk}(2018)}]{Banerjee2018}
{Banerjee}, S. \& {Kritsuk}, A.~G. 2018, \pre, 97, 023107

\bibitem[{{Benzi} {et~al.}(2010){Benzi}, {Biferale}, {Fisher}, {Lamb}, \&
  {Toschi}}]{Benzi2010}
{Benzi}, R., {Biferale}, L., {Fisher}, R., {Lamb}, D.~Q., \& {Toschi}, F. 2010,
  Journal of Fluid Mechanics, 653, 221

\bibitem[{{Benzi} {et~al.}(1993){Benzi}, {Ciliberto}, {Tripiccione}, {Baudet},
  {Massaioli}, \& {Succi}}]{Benzi1993}
{Benzi}, R., {Ciliberto}, S., {Tripiccione}, R., {et~al.} 1993, PhysRevE, 48,
  R29

\bibitem[{{Boldyrev}(2002)}]{Boldyrev2002}
{Boldyrev}, S. 2002, \apj, 569, 841

\bibitem[{{Boneberg} {et~al.}(2015){Boneberg}, {Dale}, {Girichidis}, \&
  {Ercolano}}]{Boneberg2015}
{Boneberg}, D.~M., {Dale}, J.~E., {Girichidis}, P., \& {Ercolano}, B. 2015,
  \mnras, 447, 1341

\bibitem[{{Brunt} \& {Heyer}(2013)}]{Brunt2013}
{Brunt}, C.~M. \& {Heyer}, M.~H. 2013, \mnras, 433, 117

\bibitem[{{Brunt} {et~al.}(2009){Brunt}, {Heyer}, \& {Mac Low}}]{Brunt2009}
{Brunt}, C.~M., {Heyer}, M.~H., \& {Mac Low}, M.-M. 2009, \aap, 504, 883

\bibitem[{{Burkhart} {et~al.}(2015){Burkhart}, {Collins}, \&
  {Lazarian}}]{Burkhart2015}
{Burkhart}, B., {Collins}, D.~C., \& {Lazarian}, A. 2015, \apj, 808, 48

\bibitem[{{Chira} {et~al.}(2018{\natexlab{a}}){Chira},
  {Ib{\'a}{\~n}ez-Mej{\'{\i}}a}, {Mac Low}, \& {Henning}}]{Chira2018b}
{Chira}, R.-A., {Ib{\'a}{\~n}ez-Mej{\'{\i}}a}, J.~C., {Mac Low}, M.-M., \&
  {Henning}, T. 2018{\natexlab{a}}, How do velocity structure functions trace
  gas dynamics in simulated molecular clouds?, Digital Repository (New York:
  American Museum of Natural History), doi:10.5531/sd.astro.3

\bibitem[{{Chira} {et~al.}(2018{\natexlab{b}}){Chira}, {Kainulainen},
  {Ib{\'a}{\~n}ez-Mej{\'{\i}}a}, {Henning}, \& {Mac Low}}]{Chira2018}
{Chira}, R.-A., {Kainulainen}, J., {Ib{\'a}{\~n}ez-Mej{\'{\i}}a}, J.~C.,
  {Henning}, T., \& {Mac Low}, M.-M. 2018{\natexlab{b}}, \aap, 610, A62

\bibitem[{{Dekel} \& {Krumholz}(2013)}]{Dekel2013}
{Dekel}, A. \& {Krumholz}, M.~R. 2013, \mnras, 432, 455

\bibitem[{{Elmegreen}(1993)}]{Elmegreen1993}
{Elmegreen}, B.~G. 1993, in Protostars and Planets III, ed. E.~H. {Levy} \&
  J.~I. {Lunine}, 97--124

\bibitem[{{Elmegreen}(2007)}]{Elmegreen2007}
{Elmegreen}, B.~G. 2007, \apj, 668, 1064

\bibitem[{{Falgarone} {et~al.}(2009){Falgarone}, {Pety}, \&
  {Hily-Blant}}]{Falgarone2009}
{Falgarone}, E., {Pety}, J., \& {Hily-Blant}, P. 2009, \aap, 507, 355

\bibitem[{{Federrath} {et~al.}(2011){Federrath}, {Sur}, {Schleicher},
  {Banerjee}, \& {Klessen}}]{Federrath2011}
{Federrath}, C., {Sur}, S., {Schleicher}, D. R.~G., {Banerjee}, R., \&
  {Klessen}, R.~S. 2011, \apj, 731, 62

\bibitem[{{Fleck}(1980)}]{Fleck1980}
{Fleck}, Jr., R.~C. 1980, \apj, 242, 1019

\bibitem[{{Fryxell} {et~al.}(2000){Fryxell}, {Olson}, {Ricker}, {Timmes},
  {Zingale}, {Lamb}, {MacNeice}, {Rosner}, {Truran}, \& {Tufo}}]{Fryxell2000}
{Fryxell}, B., {Olson}, K., {Ricker}, P., {et~al.} 2000, \apjs, 131, 273

\bibitem[{{Galtier} \& {Banerjee}(2011)}]{Galtier2011}
{Galtier}, S. \& {Banerjee}, S. 2011, Physical Review Letters, 107, 134501

\bibitem[{{Gnedin}(2015)}]{Gnedin2015}
{Gnedin}, O. 2015, IAU General Assembly, 22, 2256326

\bibitem[{{Goodman} {et~al.}(1998){Goodman}, {Barranco}, {Wilner}, \&
  {Heyer}}]{Goodman1998}
{Goodman}, A.~A., {Barranco}, J.~A., {Wilner}, D.~J., \& {Heyer}, M.~H. 1998,
  \apj, 504, 223

\bibitem[{{Gotoh} {et~al.}(2002){Gotoh}, {Fukayama}, \& {Nakano}}]{Gotoh2002}
{Gotoh}, T., {Fukayama}, D., \& {Nakano}, T. 2002, Physics of Fluids, 14, 1065

\bibitem[{{Hartmann} {et~al.}(2012){Hartmann}, {Ballesteros-Paredes}, \&
  {Heitsch}}]{Hartmann2012}
{Hartmann}, L., {Ballesteros-Paredes}, J., \& {Heitsch}, F. 2012, \mnras, 420,
  1457

\bibitem[{{Heyer} \& {Dame}(2015)}]{Heyer2015}
{Heyer}, M. \& {Dame}, T.~M. 2015, \araa, 53, 583

\bibitem[{{Heyer} {et~al.}(2009){Heyer}, {Krawczyk}, {Duval}, \&
  {Jackson}}]{Heyer2009}
{Heyer}, M., {Krawczyk}, C., {Duval}, J., \& {Jackson}, J.~M. 2009, \apj, 699,
  1092

\bibitem[{{Heyer} \& {Brunt}(2007)}]{Heyer2007}
{Heyer}, M.~H. \& {Brunt}, C. 2007, in IAU Symposium, Vol. 237, Triggered Star
  Formation in a Turbulent ISM, ed. B.~G. {Elmegreen} \& J.~{Palous}, 9--16

\bibitem[{{Heyer} \& {Brunt}(2004)}]{Heyer2004}
{Heyer}, M.~H. \& {Brunt}, C.~M. 2004, \apjl, 615, L45

\bibitem[{{Ib{\'a}{\~n}ez-Mej{\'{\i}}a}
  {et~al.}(2016){Ib{\'a}{\~n}ez-Mej{\'{\i}}a}, {Mac Low}, {Klessen}, \&
  {Baczynski}}]{IbanezMejia2016}
{Ib{\'a}{\~n}ez-Mej{\'{\i}}a}, J.~C., {Mac Low}, M.-M., {Klessen}, R.~S., \&
  {Baczynski}, C. 2016, \apj, 824, 41

\bibitem[{{Ib{\'a}{\~n}ez-Mej{\'{\i}}a}
  {et~al.}(2017){Ib{\'a}{\~n}ez-Mej{\'{\i}}a}, {Mac Low}, {Klessen}, \&
  {Baczynski}}]{IbanezMejia2017}
{Ib{\'a}{\~n}ez-Mej{\'{\i}}a}, J.~C., {Mac Low}, M.-M., {Klessen}, R.~S., \&
  {Baczynski}, C. 2017, \apj, 850, 62

\bibitem[{{Kolmogorov}(1941)}]{Kolmogorov1941}
{Kolmogorov}, A. 1941, Akademiia Nauk SSSR Doklady, 30, 301

\bibitem[{{Kritsuk} {et~al.}(2013){Kritsuk}, {Lee}, \& {Norman}}]{Kritsuk2013}
{Kritsuk}, A.~G., {Lee}, C.~T., \& {Norman}, M.~L. 2013, \mnras, 436, 3247

\bibitem[{Kritsuk {et~al.}(2013)Kritsuk, Wagner, \& Norman}]{Kritsuk2013a}
Kritsuk, A.~G., Wagner, R., \& Norman, M.~L. 2013, Journal of Fluid Mechanics,
  729, R1

\bibitem[{{Kritsuk} {et~al.}(2015){Kritsuk}, {Wagner}, \&
  {Norman}}]{Kritsuk2015}
{Kritsuk}, A.~G., {Wagner}, R., \& {Norman}, M.~L. 2015, in Astronomical
  Society of the Pacific Conference Series, Vol. 498, Numerical Modeling of
  Space Plasma Flows ASTRONUM-2014, ed. N.~V. {Pogorelov}, E.~{Audit}, \& G.~P.
  {Zank}, 16

\bibitem[{{Krumholz} {et~al.}(2014){Krumholz}, {Bate}, {Arce}, {Dale},
  {Gutermuth}, {Klein}, {Li}, {Nakamura}, \& {Zhang}}]{Krumholz2014}
{Krumholz}, M.~R., {Bate}, M.~R., {Arce}, H.~G., {et~al.} 2014, Protostars and
  Planets VI, 243

\bibitem[{{Larson}(1981)}]{Larson1981}
{Larson}, R.~B. 1981, \mnras, 194, 809

\bibitem[{{Mac Low}(2003)}]{MacLow2003}
{Mac Low}, M.-M. 2003, in Lecture Notes in Physics, Berlin Springer Verlag,
  Vol. 614, Turbulence and Magnetic Fields in Astrophysics, ed. E.~{Falgarone}
  \& T.~{Passot}, 182--212

\bibitem[{{Mac Low} \& {Klessen}(2004)}]{MacLow2004}
{Mac Low}, M.-M. \& {Klessen}, R.~S. 2004, Reviews of Modern Physics, 76, 125

\bibitem[{{McKee} \& {Zweibel}(1992)}]{McKee1992}
{McKee}, C.~F. \& {Zweibel}, E.~G. 1992, \apj, 399, 551

\bibitem[{{Miesch} \& {Bally}(1994)}]{Miesch1994}
{Miesch}, M.~S. \& {Bally}, J. 1994, \apj, 429, 645

\bibitem[{{Miyamoto} {et~al.}(2014){Miyamoto}, {Nakai}, \&
  {Kuno}}]{Miyamoto2014}
{Miyamoto}, Y., {Nakai}, N., \& {Kuno}, N. 2014, \pasj, 66, 36

\bibitem[{{Padoan} {et~al.}(2003){Padoan}, {Boldyrev}, {Langer}, \&
  {Nordlund}}]{Padoan2003}
{Padoan}, P., {Boldyrev}, S., {Langer}, W., \& {Nordlund}, {\AA}. 2003, \apj,
  583, 308

\bibitem[{{Padoan} {et~al.}(2006){Padoan}, {Juvela}, {Kritsuk}, \&
  {Norman}}]{Padoan2006}
{Padoan}, P., {Juvela}, M., {Kritsuk}, A., \& {Norman}, M.~L. 2006, \apjl, 653,
  L125

\bibitem[{{Padoan} {et~al.}(2016{\natexlab{a}}){Padoan}, {Pan}, {Haugb{\o}lle},
  \& {Nordlund}}]{Padoan2016}
{Padoan}, P., {Pan}, L., {Haugb{\o}lle}, T., \& {Nordlund}, {\AA}.
  2016{\natexlab{a}}, \apj, 822, 11

\bibitem[{{Padoan} {et~al.}(2016{\natexlab{b}}){Padoan}, {Pan}, {Haugb{\o}lle},
  \& {Nordlund}}]{Padoan2016a}
{Padoan}, P., {Pan}, L., {Haugb{\o}lle}, T., \& {Nordlund}, {\AA}.
  2016{\natexlab{b}}, \apj, 822, 11

\bibitem[{{Roman-Duval} {et~al.}(2011){Roman-Duval}, {Federrath}, {Brunt},
  {Heyer}, {Jackson}, \& {Klessen}}]{RomanDuval2011}
{Roman-Duval}, J., {Federrath}, C., {Brunt}, C., {et~al.} 2011, \apj, 740, 120

\bibitem[{{Schmidt} {et~al.}(2008){Schmidt}, {Federrath}, \&
  {Klessen}}]{Schmidt2008}
{Schmidt}, W., {Federrath}, C., \& {Klessen}, R. 2008, Physical Review Letters,
  101, 194505

\bibitem[{{Seifried} {et~al.}(2017){Seifried}, {Walch}, {Girichidis}, {Naab},
  {W{\"u}nsch}, {Klessen}, {Glover}, {Peters}, \& {Clark}}]{Seifried2017b}
{Seifried}, D., {Walch}, S., {Girichidis}, P., {et~al.} 2017, ArXiv e-prints

\bibitem[{{She} \& {L\'{e}v\^{e}que}(1994)}]{She1994}
{She}, Z.-S. \& {L\'{e}v\^{e}que}, E. 1994, Physical Review Letters, 72, 336

\bibitem[{{Solomon} {et~al.}(1987){Solomon}, {Rivolo}, {Barrett}, \&
  {Yahil}}]{Solomon1987}
{Solomon}, P.~M., {Rivolo}, A.~R., {Barrett}, J., \& {Yahil}, A. 1987, \apj,
  319, 730

\bibitem[{{Tan} {et~al.}(2013){Tan}, {Shaske}, \& {Van Loo}}]{Tan2013}
{Tan}, J.~C., {Shaske}, S.~N., \& {Van Loo}, S. 2013, in IAU Symposium, Vol.
  292, Molecular Gas, Dust, and Star Formation in Galaxies, ed. T.~{Wong} \&
  J.~{Ott}, 19--28

\bibitem[{{Truelove} {et~al.}(1998){Truelove}, {Klein}, {McKee}, {Holliman},
  {Howell}, {Greenough}, \& {Woods}}]{Truelove1998}
{Truelove}, J.~K., {Klein}, R.~I., {McKee}, C.~F., {et~al.} 1998, \apj, 495,
  821

\bibitem[{{Turk} {et~al.}(2012){Turk}, {Oishi}, {Abel}, \& {Bryan}}]{Turk2012}
{Turk}, M.~J., {Oishi}, J.~S., {Abel}, T., \& {Bryan}, G.~L. 2012, \apj, 745,
  154

\bibitem[{{V{\'a}zquez-Semadeni} {et~al.}(2006){V{\'a}zquez-Semadeni}, {Ryu},
  {Passot}, {Gonz{\'a}lez}, \& {Gazol}}]{Vazquez2006}
{V{\'a}zquez-Semadeni}, E., {Ryu}, D., {Passot}, T., {Gonz{\'a}lez}, R.~F., \&
  {Gazol}, A. 2006, \apj, 643, 245

\bibitem[{{Zernickel}(2015)}]{Zernickel2015}
{Zernickel}, A. 2015, PhD thesis, I.~Physikalisches Institut der
  Universit{\"a}t zu K{\"o}ln, Z{\"u}lpicher Stra{\ss}e 77, 50937, K{\"o}ln,
  Germany

\end{thebibliography}

\appendix

\section{Computation and Fitting Procedures}\label{appFitting}

In this section we provide more details on how we compute the VSFs and their scaling parameters.
As described in Sect.~\ref{methods:vsf} the discussions in the main part of this manuscript are based on VSFs that are computed from average relative velocities. 
This means the following:

We map the 3D FLASH adaptive mesh refinement data of the original simulations \citepalias{IbanezMejia2016,IbanezMejia2017} onto uniform grid data cubes of 40~pc on a side with 0.1~pc zones centred on the centres of the molecular clouds. 

The FLASH simulations take ten levels of resolution into account; and the decision which region is resolved up to which level depends on the density of the respective grid cell \citepalias{IbanezMejia2017}.
Therefore, the original simulations cover a range of resolutions from 30.4~pc in regions of $|z| > 10$~kpc above and below the galactic midplane down to 0.06--0.10~pc within (100~pc)$^3$ boxes around the centre of mass of the three molecular clouds presented in this manuscript, as well as in \citetalias{IbanezMejia2017} and \citetalias{Chira2018}. 
Therefore, the resolution we have chosen for the data cubes corresponds to either the highest resolved level offered by the simulations (in the case of \texttt{M8}) or a slight coarsening of the most resolved level (in the cases of \texttt{M3} and \texttt{M4}).
The zoom-in regions modelled at the highest resolution level extend well beyond the volume we use in our analysis.
We have adopted this size as it is sufficient for covering the clouds' volumes, and thus more efficient for further analyses. In addition, this choice protects us from resolution edge effects as the transitions from this resolution level to the next lowest are several tens of zones away. 

No matter whether a density threshold is applied or not, the (40 pc)$^3$ cubes still include too much data to compute complete velocity structure functions including all lags from all points.
Therefore, to derive the VSFs we coarsen the grid of projected lag distances, $\ell_i = |\vec{\ell}|_i$ to cover the range 0.8--30~pc with only 40 equidistant bins; relative to each of the starting coordinates of our samples. 

After mapping the data onto the uniform grid, we apply two approaches: 
The first considers only zones above the density threshold, $n_\mathrm{cloud}$.
These zones represent the starting points $\vec{x}$ (see Eqs.~\ref{equ:method:def_vsf}--\ref{equ:method:def_vsf_1d}).
Our routine calculates the lag distances between these zones and every other zone in the sample, as well as the relative velocities of the gas within the given zones. 
The individual lag distances are binned using spherical shells around the starting zones that range from inner radii $\ell_{i}$ to outer radii $\ell_{i+1}$. 
By doing so, we compute the discrete VSFs presented in the main part of this paper with the relative velocities and product of densities, $\rho(\vec{x}) \cdot \rho(\vec{x}+\vec{\ell})$, measured from zones within the individual shells.

The second approach targets the case when we do not apply any density threshold (i.e.~setting $n_\mathrm{cloud} =0$).
In this case, we use a random number generator (\texttt{random.rand} from numpy) to choose 5\% of the total zones, and do the analysis on them.
We emphasise that this does not mean that we only calculate the relative velocities between these zones. Rather this subsample of zones represent the starting vectors $\vec{x}$ to which the velocities of all other zones $\vec{x} + \vec{\ell}$ in the same cube are compared to. This way we reduce the risk of ignoring or emphasising any spatial direction or angle.
As it is too computationally expensive to derive all relative velocities between all zones within discrete shells, as we have done in the first approach, we derived a discrete distribution of relative velocities as function of lag distance using a discrete fast Fourier transform (\texttt{FFT}). 
These distributions are based on the same grid of lag distances we have already utilised for the spherical shells above.
Therefore, we can use the results of the \texttt{FFT} in the same way as the results from the first approach to derive the VSFs.

With both approaches we obtain a set of discrete descriptions of VSFs as a function of  lag  and time (by computing VSFs for successive code outputs).
In order to derive the scaling parameters $\zeta$ of the VSFs as a function of time and order, we fit the power-law relation presented in Eq.~(\ref{equ:method:fitting}) to the measured VSFs, using the python \texttt{curve\_fit} package.
Due to the rather irregular behaviour of our VSFs at larger scales, we define the weighting function as
\begin{equation}
w\left(\ell\right) = \begin{cases}
    0 & \ell \leq~\mathrm{0.8~pc} \\
    1 & \mathrm{0.8~pc} \leq \ell \leq~\mathrm{8~pc} \\
    \mathrm{1~pc}\,\ell^{-1} & \ell >~\mathrm{8~pc}.
\end{cases}
\end{equation}
\noindent Although the average radii of our clouds are, on average, larger than 8~pc we choose this limit due to the variable behaviour of the VSFs at scales of that size and larger.
Investigation of different weightings would be a fruitful topic of further investigation.

As can be seen in Fig.~\ref{pic:results:vsf_example}, as well as in the figures shown in Appendix~\ref{appInertial}, the shape of VSFs changes over time as different forces act, as we explain in Sect.~\ref{results}. 

The similarity parameters, $Z$, are computed by applying Eq.~\ref{equ:method:z_def} on the results of the fitting procedure, and results are presented in Sect.~\ref{results}, as well as in Appendix~\ref{appInertial}.

\begin{figure*}
\centering
\includegraphics[width=0.9\textwidth]{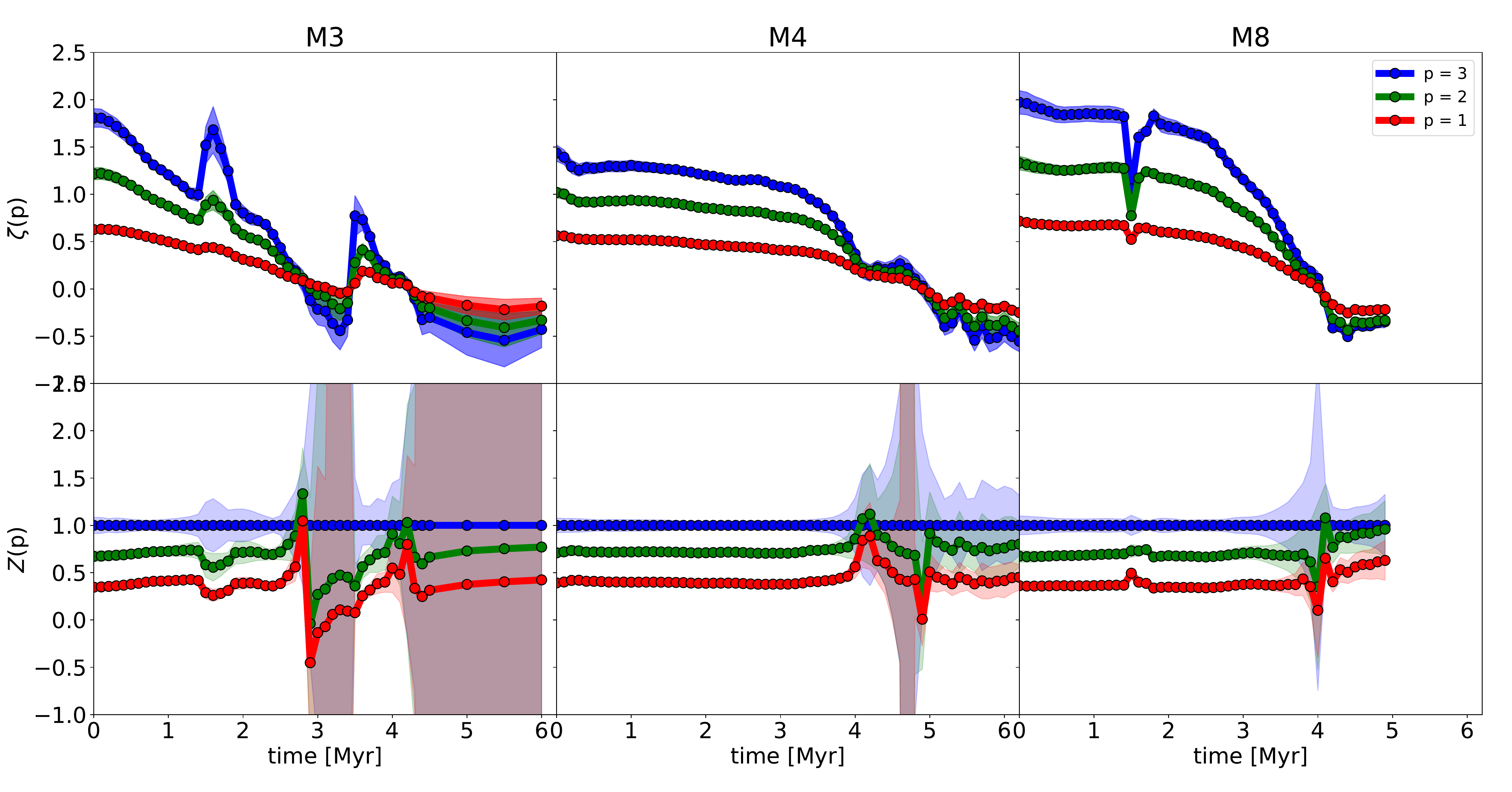}
\caption{
    Data presented in Fig.~\ref{pic:results:zeta_all}a, with
    shaded areas behind the data representing the error ranges of the computed $\zeta$ (\textit{top}) and $Z$ (\textit{bottom}).
}
\label{pic:appFitting:error_vsfhr04_zeta_z}
\end{figure*}

The fit also provides the $\chi^2$ errors for the measured values of $\zeta$. 
In Fig.~\ref{pic:appFitting:error_vsfhr04_zeta_z} we show a reduced version of Fig.~\ref{pic:results:zeta_all}(a), where we only plot the time evolution of $\zeta$ for all three clouds, along with shades of the same colours of the respective lines that represent the errors.
We see that the relative errors, $\Delta \zeta / \zeta$ mostly remain within a range of 5--12\%. 
The errors of $Z$ are computed by Gaussian error propagation
\begin{align}\Delta Z(p) &= \sqrt{ \left( \frac{\partial Z(p)}{\partial \zeta(p)} \cdot \Delta\zeta(p) \right)^2 + \left( \frac{\partial Z(p)}{\partial \zeta(3)} \cdot \Delta\zeta(3) \right)^2 } \\
    &= \sqrt{ \left( \frac{\Delta\zeta(p)}{\zeta(3)} \right)^2 + \left( \frac{ \zeta(p) \cdot \Delta\zeta(3)}{\zeta(3)^2} \right)^2 }.
    \label{equ:appFitting:z_error}
\end{align}
\noindent In general, the relative errors of $Z$ are, as well, around 10\%, though we do see exceptions with very large errors. 
The reason for these is that at these times $\zeta$(3) approaches zero, causing Eq.~(\ref{equ:appFitting:z_error}) to diverge.


\section{Inertial Range}\label{appInertial}

For our analysis it is essential to verify that there is a reasonable range of scales $\ell$ below the driving range within which the simulations are not dominated by numerical effects, so that they resolve the inertial range of any turbulent cascade.  
We note that the fitting region for our VSFs starts at lags greater than eight zones, ensuring that the numerical dissipation range lies at smaller scales.
To understand the non-power law behaviour that we nonetheless find, in the following subsections we offer more examples that show VSFs of all three clouds at different times and considering our different analysis approaches. 
We focus on the standard analysis in Sect.~\ref{Bsub:standard}, on the
analysis neglecting the density threshold in Sect.~\ref{Bsub:full},
and on the impact of varying the resolution in collapsing regions in Sect.~\ref{Bsub:Jeans}.

\begin{figure*}
    \centering
    \includegraphics[width=\textwidth]{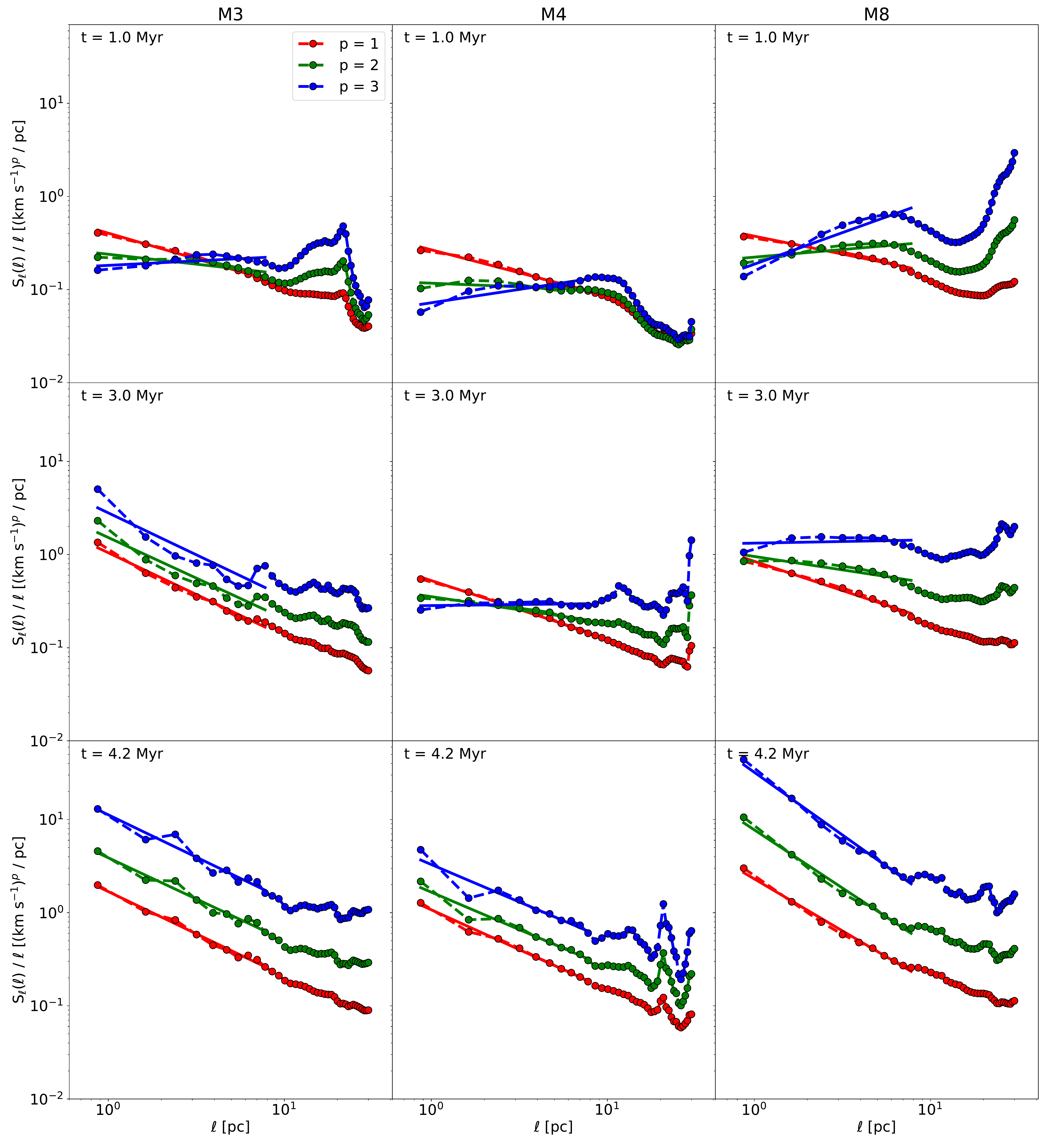}
    \caption{
        As Fig.~\ref{pic:results:vsf_example}, but plotting the relation between S$_{\ell}$ / $\ell$ as function of lag scale $l$ and order $p$.
    }
    \label{pic:appInertial:examples_with_threshold_sl_vs_l}
\end{figure*}

\begin{figure*}
    \centering
    \includegraphics[width=\textwidth]{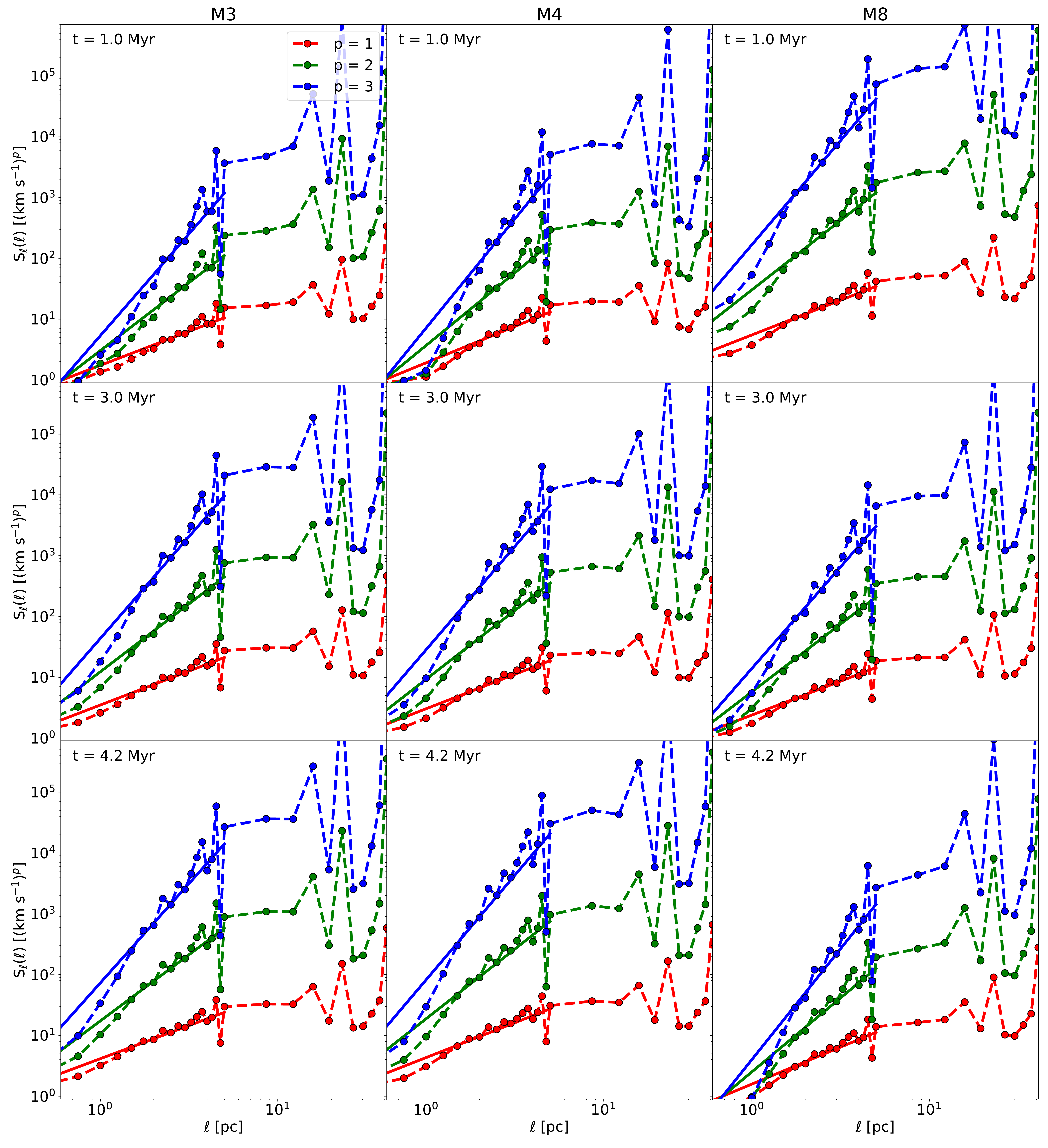}
    \caption{
        As Fig.~\ref{pic:results:vsf_example}, but based on data without density threshold.
    }
    \label{pic:appInertial:examples_without_threshold_s_vs_l}
\end{figure*}

\begin{figure*}
    \centering
    \includegraphics[width=\textwidth]{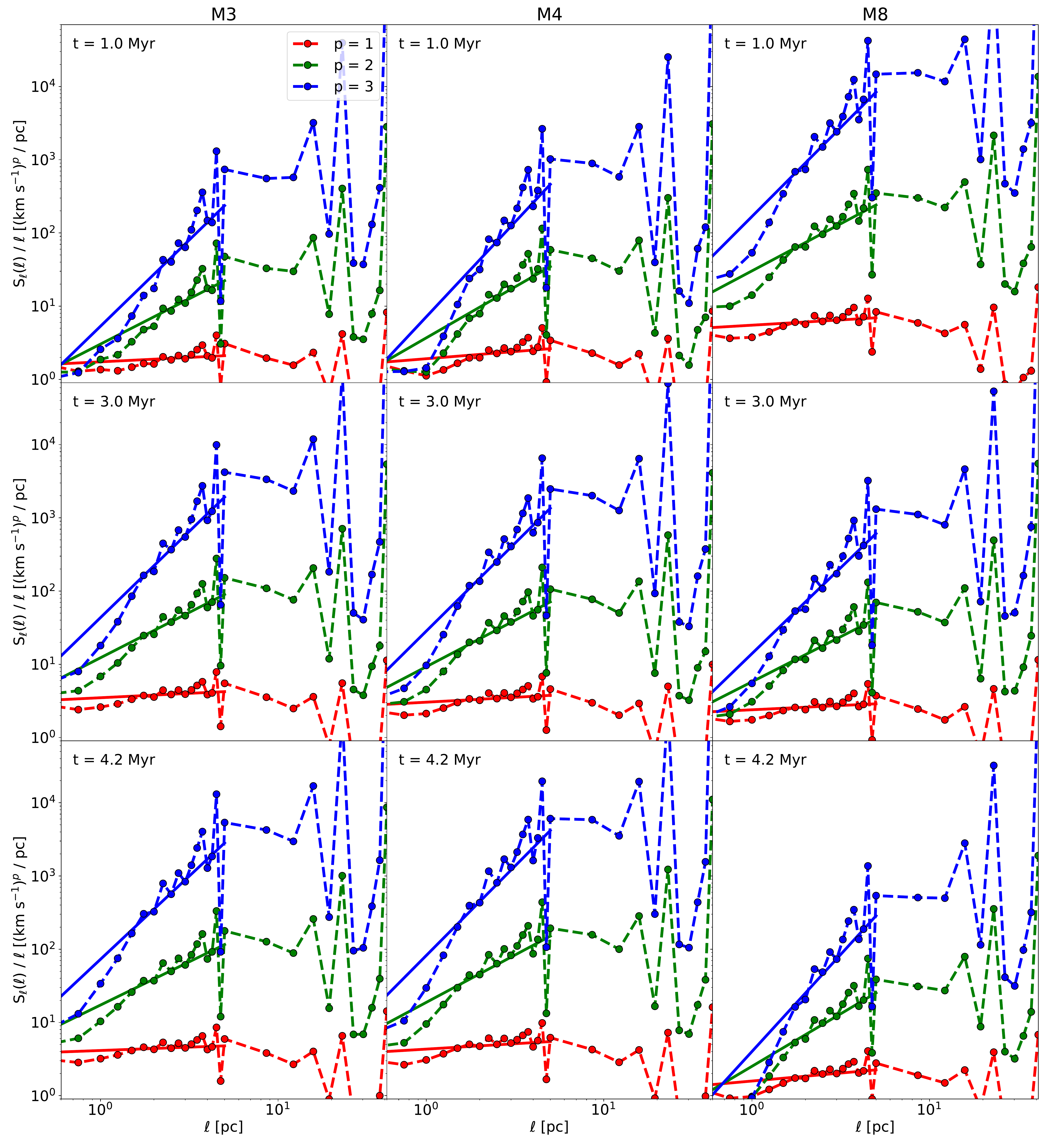}
    \caption{
        As Fig.~\ref{pic:appInertial:examples_with_threshold_sl_vs_l}, but based on data without density threshold.
    }
    \label{pic:appInertial:examples_without_threshold_sl_vs_l}
\end{figure*}

\begin{figure*}
    \centering
    \includegraphics[width=\textwidth]{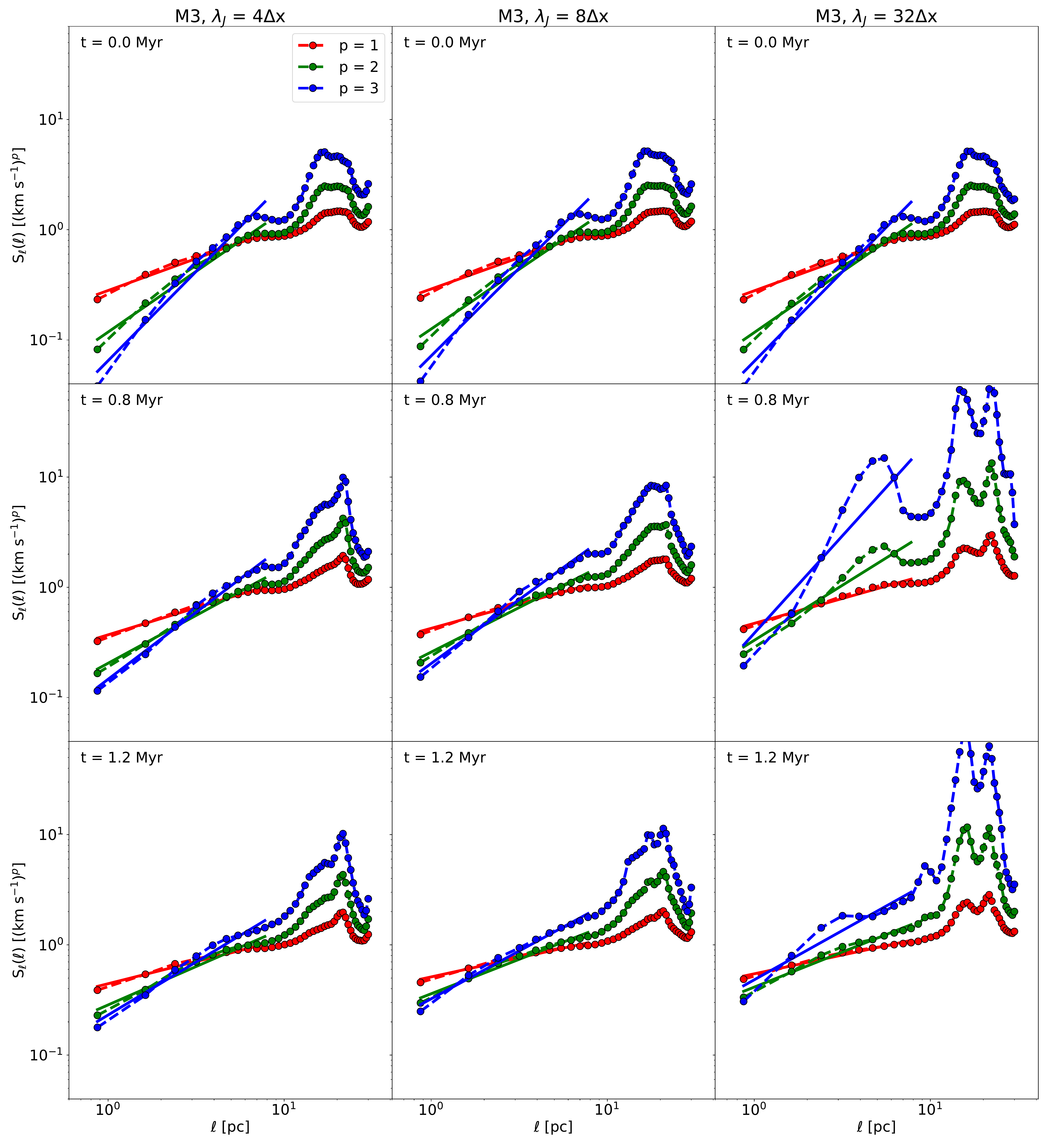}
    \caption{
        The figures show additional examples of VSFs, based on data of \texttt{M3} with density threshold, at the different refinement levels (\textit{left} to \textit{right}) $\lambda$~=~4~$\Delta$x, $\lambda$~=~8~$\Delta$x, and $\lambda$~=~8~$\Delta$x as function of lag scale $\ell$ and order $p$. 
        The examples are given for three different time steps, namely (\textit{top} to \textit{bottom}) t~=~0.0~Myr, 0.8~Myr, and 1.2~Myr.
        The dots (connected by dashed lines) show the values computed from the simulations. 
        The solid lines represent the power-law relation fitted to the respective structure functions.
    }
    \label{pic:appInertial:examples_jeans_s_vs_l}
\end{figure*}

\begin{figure*}
    \centering
    \includegraphics[width=\textwidth]{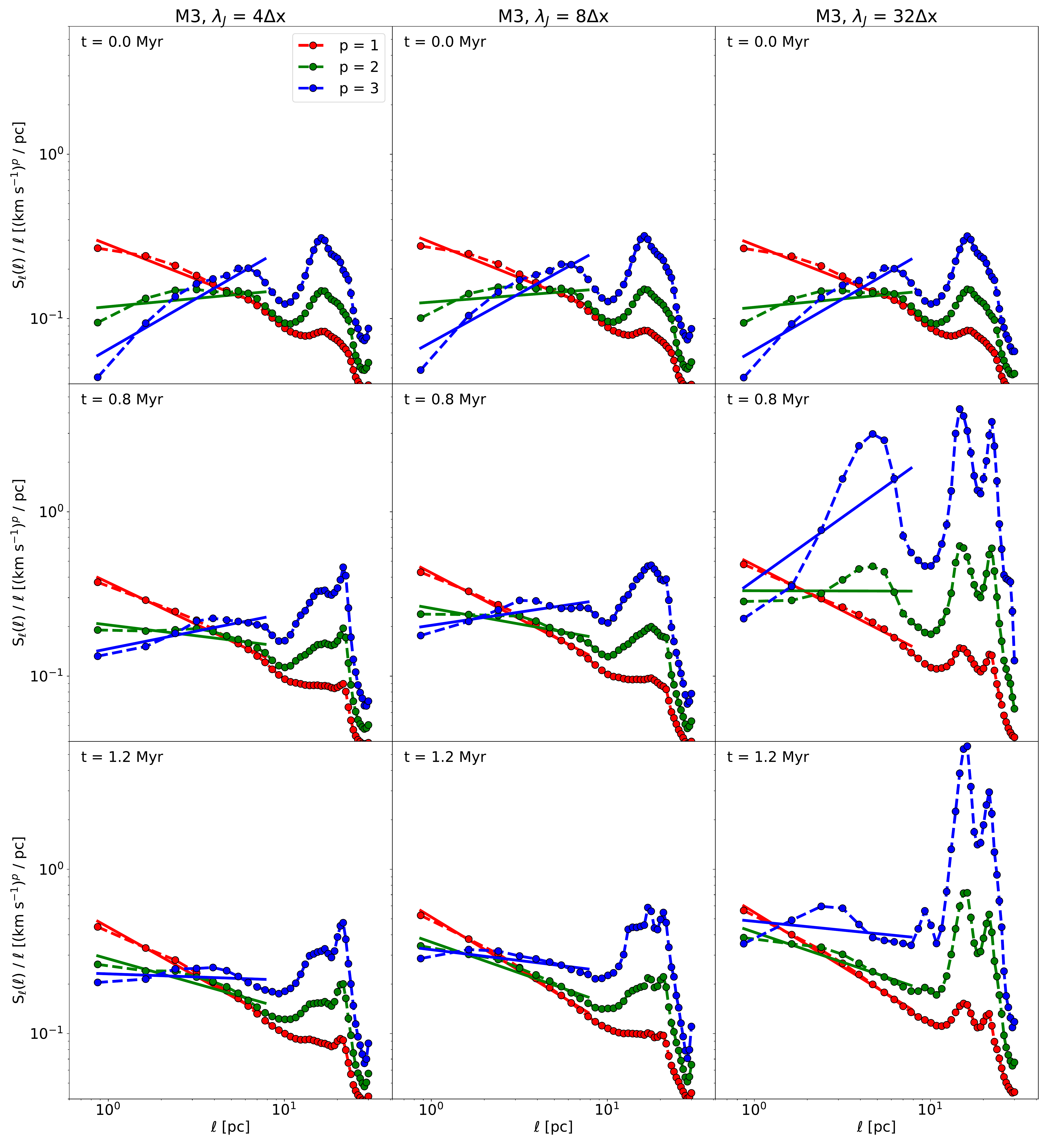}
    \caption{
        As Fig.~\ref{pic:appInertial:examples_jeans_s_vs_l}, but plotting the relation between S$_{\ell}$ / $\ell$ as function of lag scale $l$ and order $p$.
    }
    \label{pic:appInertial:examples_jeans_sl_vs_l}
\end{figure*}

\subsection{Standard Analysis}\label{Bsub:standard}

Fig.~\ref{pic:appInertial:examples_with_threshold_sl_vs_l} extends the data presented in Sect.~\ref{results:normal} and discussed in Sect.~\ref{discussion:normal}.
The figure uses the same format as Fig.~\ref{pic:results:vsf_example}.
The straight lines within the plots indicate the power-law relation that we have fitted onto the VSFs, considering the range 0.8~$\leq\,\ell\,\leq$~8~pc.

We see that in most of the cases the VSFs are in good agreement with the described power-law relation within the fitted ranges (e.g., Fig.~\ref{pic:results:vsf_example}). 
However, there are cases when the VSF is not well reproduced by a simple, single power-law function, such as \texttt{M3} at $t=3.0$~Myr in Fig.~\ref{pic:results:vsf_example}.
This appears to occur particularly when the dominant mechanism driving turbulence changes from large-scale driving to internal contraction. 
At this point the  larger scales of the clouds (larger~$\ell$) are still dominated by external driving, while the smaller scales (smaller~$\ell$) start to accelerate mostly due to gravitational fragmentation and infall motions.
The consequence is that the actual VSF is a superposition of two processes that amplify the relative motions of the gas differently and on different scales.
A single power-law cannot describe this scenario adequately.

\subsection{No Density Threshold} \label{Bsub:full}

Figs.~\ref{pic:appInertial:examples_without_threshold_s_vs_l} and~\ref{pic:appInertial:examples_without_threshold_sl_vs_l} extend the set of VSFs for the box without a density threshold presented in Sect.~\ref{results:densthres} and discussed in Sect.~\ref{discussion:densthres}.

Here we also find problems describing the behaviour of small-scale motions with $\ell\,\lesssim$~2~pc.
In this case we primarily capture the turbulent motions within the low-density ISM. 
Contrary to the modelled MCs, the ISM is not organised in hierarchical structures, and the turbulence within the ISM is predominantly driven by SN explosions.
These produce structured correlations at the large scales of entire blast waves as well as the small scales across shock fronts.
Yet, we see that we can fit the VSFs with a power law within the intermediate lag scale ranges (2~pc~$\lesssim\,\ell\,\lesssim$~8~pc), suggesting this range of length scales is dominated by a turbulent cascade.

\subsection{Varying Refinement} \label{Bsub:Jeans}

Figs.~\ref{pic:appInertial:examples_jeans_s_vs_l} and~\ref{pic:appInertial:examples_jeans_sl_vs_l} extend the set of VSFs presented for clouds with varying amounts of refinement in gravitationally collapsing regions in Sect.~\ref{results:refinement} and discussed in Sect.~\ref{discussion:refinement}.

A final example of VSFs not following a power law at intermediate scales is given by cloud \texttt{M3} at $t=0.8$~Myr and with $\lambda_\mathrm{J} = 32\Delta x$ in Fig.~\ref{pic:appInertial:examples_jeans_s_vs_l}.
This represents the VSF of a cloud that is interacting with a SN blast wave.
In this case the maximal amplification is neither at the scale of the cloud, where the turbulence is driven by external sources, nor on small scales where gravitational contraction acts.
Instead the local maximum is the intermediate scale. 
Considering the morphology of the cloud and the cloud's environment this can only occur when the shock front of a SN is currently propagating through the cloud. 
Thus, the VSF here is a superposition of three driving mechanisms:
first, the external large-scale driving; second, self-gravity that leads to contractive motions; and finally the shock jump that injects kinetic energy as it moves through the cloud. 
The effect of the shock, however, is only local and short-lived as the injected turbulence decays quickly (compare with \texttt{M3} at $t=1.2$~Myr and with $\lambda_\mathrm{J} = 32\Delta x$ in Fig.~\ref{pic:appInertial:examples_jeans_s_vs_l}).

\end{document}